\documentclass{aa}
\usepackage{txfonts}
\usepackage{graphics}
\usepackage{latexsym}
\usepackage{natbib}

\begin{document} 
\title{Gravitational scattering of stars and clusters and the heating of the Galactic disk}

\titlerunning{High altitude clusters and the heating of the Galactic disk}

\author{
        Bengt Gustafsson\inst{1,3}
   \and Ross P. Church\inst{2}
   \and Melvyn B. Davies\inst{2}
    \and Hans Rickman\inst{1,4}
       }

\authorrunning{Gustafsson et al.}

\institute{Department of Physics and Astronomy, Uppsala University, Box 515, S-751\,20 Uppsala, Sweden
     \and  Department of Astronomy and Theoretical Physics, Lund Observatory, Box 43,
         SE-221 00, Lund, Swedem
     \and  NORDITA, Roslagstullsbacken 23, SE-106\,91 Stockholm, Sweden
     \and PAS Space Research Center, Bartycka 18A, PL-00-716 Warsawa, Poland }

\date{Received <date> / Accepted <date>}
\offprints{Bengt Gustafsson,
\email{Bengt.Gustafsson@physics.uu.se}}

 \abstract
{
Could the velocity spread, increasing with time, in the Galactic disk be explained
as a result of gravitational interactions of stars with giant molecular clouds (GMCs) and spiral arms?
Do the old open clusters high above the Galactic plane provide clues to this question? 
}
{
We explore the effects on stellar orbits of scattering by inhomogeneities in the Galactic potential
due to GMCs, spiral arms and the Galactic bar, and whether high-altitude 
clusters could have formed in orbits closer to the Galactic plane and later
been scattered. 
}
{
Simulations of test-particle motions are performed in a realistic Galactic potential. The effects of the internal
structure of GMCs are explored. The destruction of clusters in GMC collisions is treated in detail
with $N$-body simulations of the clusters.
}
{
The observed velocity dispersions of stars as a function of time
are well reproduced. The GMC structure is found to be significant, but adequate models 
produce considerable scattering effects.  
The fraction of simulated massive old open clusters, scattered into orbits with $|z|>400$ pc, is typically $0.5\%$, in agreement
with the observed number of high-altitude clusters and consistent with the 
present formation rate of massive open clusters.
} 
{
The heating of the thin Galactic disk is well explained by gravitational scattering by GMCs and spiral arms,
if the local correlation between the GMC mass and the corresponding voids in 
the gas is not very strong. Our results suggest that the high-altitude metal-rich clusters were formed 
in orbits close to the Galactic plane and
later scattered to higher orbits. It is possible, though not very probable, that the Sun formed in such a cluster 
before scattering occurred. 
}


\maketitle 


\section{Introduction}

\begin{figure}[htbp]
\centering
\resizebox{\hsize}{!}{\includegraphics{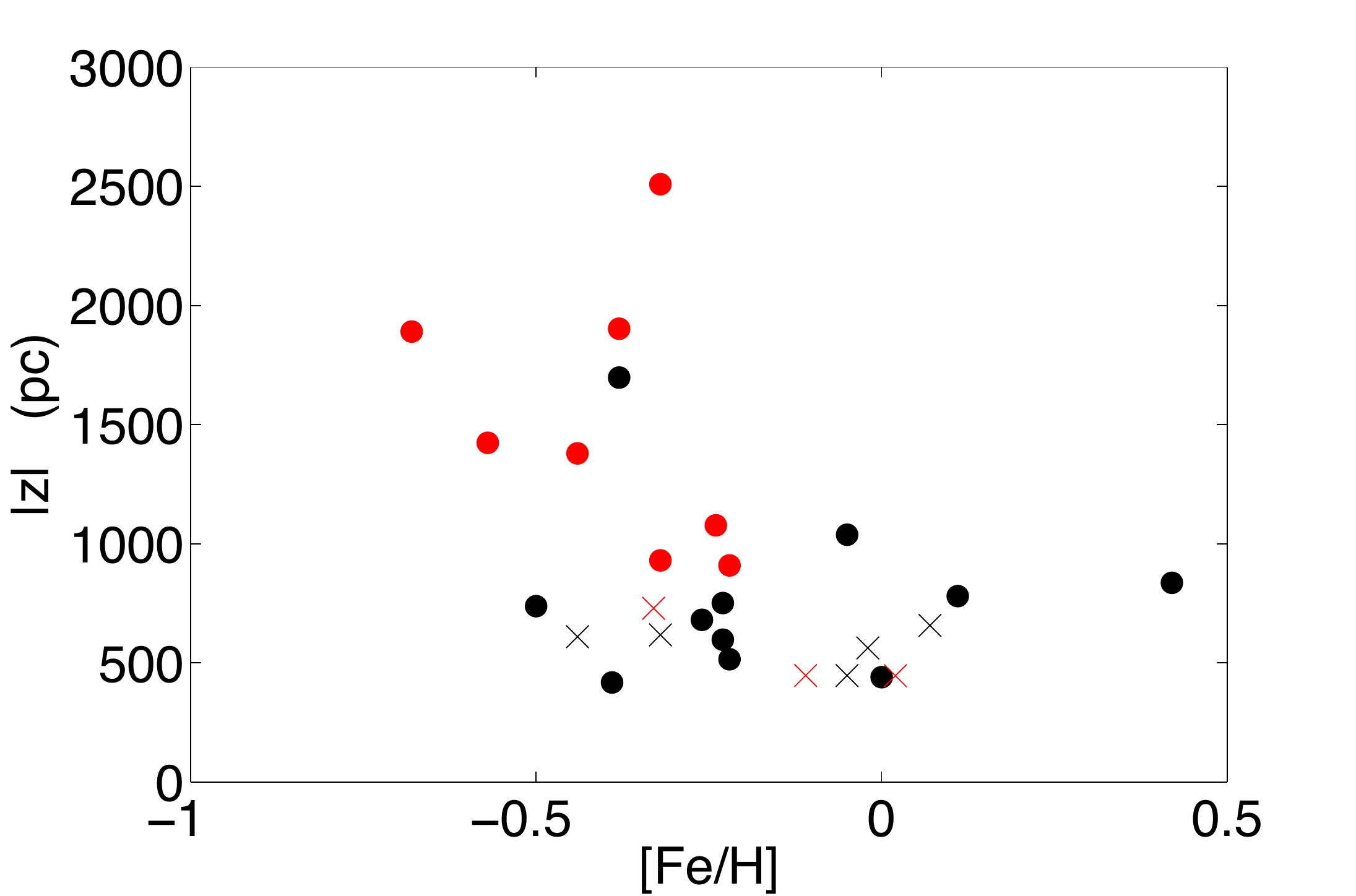}}
\caption{The distribution of high-altitude ($|z|>400$ pc) clusters with positive and negative Galactic latitudes (black and red/gray, respectively). Dots indicate clusters 
older than 1.0 Gyr while crosses correspond to lower ages. Data were taken from \citet{Heiter14}, \citet{Paunzen06},
\citet{WEBDA}. The four old clusters at highest metallicity
are marked by the four black dots to the right (in order left to right: NGC\,2420, M\,67, NGC\,188 and NGC\,6791}
\label{cluster_z}
\end{figure}

Several relatively metal-rich and massive open clusters are located high above the Galactic plane. 
From \citet{VandePutte10} we find that 8\% of all 481 open clusters in their study have  $\vert z \vert >$ 400 pc. 
The number is decreased to about 4\%  if we limit ourselves to Solar metallicities (cf their fig 7). 
Among the 78 clusters with reliable spectroscopic metallicities compiled by \citet{Heiter14}, 7 
have $|z|>400$ pc and yet close to Solar metallicity. In Fig. \ref{cluster_z} we display the known open clusters with $|z|>400$ pc. We find
4 such clusters  with [Fe/H] $>-0.1$ and an age $>$ 1.0 Gyr, M\,67, NGC 188, NGC 2420 and NGC 6791, all northern  
which could reflect incompleteness and selection effects in the data. 
Data for these clusters are listed in Table \ref{tab:cluster_data} as taken from the sources given. 

The relatively old, metal-rich and yet populous open cluster M\,67 is located
at a height $z$ above the Galactic plane of about 400-450 pc  (\citet{Sarajedini09}, \citet{VandenBerg04}, \citet{Friel95}) 
and +36 degrees in Galactic longitude from the anti-centre direction. Its orbit has presently an eccentricity of about 0.13, 
and its distance from the centre of the Galaxy, now being close to 9 kpc, is estimated to vary between 8 to 10 kpc or possibly 6 to 10 kpc, 
depending on the mass in spiral arms assumed at the orbit calculations (\citet{Pichardo12}).  
It has a metallicity very close to Solar  and an age of about 3.5 - 4.8 Gyr (see \citet{Onehag11} for references). 

NGC 188, although being more distant and probably older than M\,67, has a metallicity and mass similar to or even greater 
(\citet{Friel10}, \citet{Bonatto05}). NGC\,2420 was earlier regarded to be a transition system between solar-metallicity open
clusters and more metal-poor globular clusters. However, for this cluster more recent analyses with high-resolution spectroscopy suggest a metallicity
ranging from [Fe/H] = -0.05 to -0.20 (see \citet{Carrera13}), i.e. rather close to solar ([Fe/H]$\equiv 0.0)$. Age estimates vary between 1 and 3 Gyrs (see references in Table \ref{tab:cluster_data})
and in \citet{Carrera13}). The highly
interesting and quite old cluster NGC\,6791 (\citet{Brogaard12}) seems to be unique in showing different abundances for different stars, with an Na/O anti-correlation 
similar to that found in globular clusters, suggesting that several generations
of stars have formed while the cluster was massive enough to retain the material expelled by AGB stars  within the cluster. 
\citet{Carraro13} followed \citet{Jilkova12} in speculating that it formed in the Galactic bulge and then migrated to its
present position, 7 kpc away from the Galactic centre. 
In addition to these four clusters, we have included one similarly metal-rich, old and populous cluster at a
slightly lower altitude, NGC\,7142, into the Table. 

\begin{table}[h]
\caption{Metal-rich and old Galactic clusters at high altitudes. Data are from the references given, complemented with Netopil et al. (2012). Metallicities are from 
H14 = Heiter et al. (2014), and ages from Paunzen \& Netopil (2006)}
  \centering
  \begin{tabular}[l]{lccccl}
    \hline
   Cluster  & z &  Distance & [Fe/H] & Age & References \\
   & kpc & kpc &  & Gyr & \\
   \hline 
     M\,67  & 0.45 & 0.9 & 0.02 & 4 & \"{O}nehag \\
               &         &        &        &     &    et. al. (2014) \\
     NGC\,188 & 0.8 & 1.8 & -0.02 & 6.2 & Meibom \\
               &         &        &        &      &        et al. (2009), H14 \\
     NGC\,2420  & 1.0 & 3.1 & -0.05 & 1.1 & Netopil\\
               &         &        &        &      & et al. (2012) , H14 \\
     NGC\,6791 & 0.8 & 4.1  &  0.4  &  8.3 & Brogaard\\
              &          &        &        &      &  et al. (2012), H14\\
     NGC\,7142 & 0.38 & 2.3 & 0.11 &  3 & Strai{\v z}ys\\ 
                    &          &       &         &     &  et al. (2014), H14\\
   \hline
  \end{tabular}
  \label{tab:cluster_data}
\end{table}

One fundamental reason for studying the nature of old metal-rich clusters at high latitudes is the problem
of understanding the evolution of the Galactic disk and of galaxy disks in general. In classical papers \citet{Spitzer51} and \citet{Spitzer53} 
suggested that the gradual increase of the scatter of stellar velocities with age in the Solar neighbourhood  
is due to gravitational scattering by ``interstellar gas complexes". Accordingly, the later discovered giant molecular clouds (GMCs) became main candidates responsible for this
so-called "disk heating". Calculations by \citet{Lacey84} did, however, not reproduce the observed scatters in radial ($\sigma_U$), azimuthal ($\sigma_V$) and
perpendicular ($\sigma_W$) directions relative to the Disk; note however, that \citet{Villumsen83} obtained a better agreement with observations. 
\citet{Barbanis67}, \citet{Carlberg85}, \citet{Carlberg87} and \citet{Jenkins90} suggested that the acceleration in the plane
was due to transient spiral structure, while the scattering against the GMCs partially redirected the velocities into the W direction.  Later \citet{Ida93} and \citet{Shiidsuka99}
found that GMCs alone could, indeed give proper axis ratios for the velocity ellipsoid, see also \citet{Sellwood08} and \citet{Sellwood13}. Yet, the
effects of spiral structure, notably in the U and V velocities, and also the Galactic bar (see \citet{Saha10}, \citet{Grand15}, and \citet{Athanassoula13} and references therein)
may indeed be significant. Other mechanisms that have been suggested to play a role are 
infall of satellite galaxies and other ``cosmic sub structure"  (\citet{ Kazantzidis09}), including massive black holes (\citet{Hanninen02}, \citet{Hanninen04}) and dark-matter halos, as well as 
collective effects like buckling instabilites or bending waves in the Disk (\citet{Sotnikova03}, \citet{Saha10}, \citet{Griv97}). 
The observed and rather smooth increase of $\sigma_U$, $\sigma_V$ and
$\sigma_W$ in unison with time may speak against more dramatic irregular mechanisms (cf. \citet{Sellwood13}, see also \citet{Zasov13}). With \citet{VandeKruit11} we conclude that
there is still ``much uncertainty about the heating of the thin disk. Some of this uncertainty is due to uncertainty in the observational
relation between stellar ages and velocity dispersions, because stellar ages are so difficult to measure".  
We shall limit the present
study to the evolution of the Galactic disk from the formation of the Sun to presently, partly because the observations of heating for older Thin-Disk stars are limited and
also influenced by the mixing-in of Thick-Disk stars, probably affected by additional heating mechanisms. Morevover, the conditions in the Galactic disk, e.g. as regards
star formation and density of GMCs, are more uncertain the longer we look backwards in time, making simulations of the evolution more uncertain. 

A reason for exploring the connection between the heating of the Galactic disk and the existence of high-altitude clusters is the
possibility that the latter could illuminate the general heating mechanisms. The response of the young clusters close to the Galactic plane 
to the mechanisms, whatever they are, might not be similar to those of stars of similar ages. In particular, nearby interaction, e.g. with 
a GMC, may break up the cluster. Also, it is interesting in itself to explore whether the existence and frequency of the clusters at high latitudes could at all
be consistent with reasonable heating mechanisms for the disk in general, or whether the clusters have to be explained by other mechanisms,
such as interaction between the Disk gas and massive infalling objects like High Velocity Clouds or globular clusters, or shock interaction between 
spiral density waves and a thick magnetized Galactic disk, pushing up star-forming gas to high latitudes. For a review of such ``exotic formation scenarios", see
Appendix A.2 and \citet{VandePutte10}.  

In the present paper, some focus will be on the cluster M\,67, being the most well studied of the old metal-rich clusters, at high altitudes.
One special reason for wondering about the origin of M\,67 is its similarity in age and chemical composition with the Sun. 
In fact, \citet{Onehag11} found one Solar-twin star in the cluster to be more Solar-like than almost any known twins in the Solar neighbourhood and speculated that the
Sun might even have had an origin in the cluster. The Solar-identical abundances of the cluster were later verified by the analysis of 13 more stars in M\,67 (\citet{Onehag14}). 
The possibility of a Solar origin in the cluster was, however, refuted by \citet{Pichardo12} who argued that the kicking out 
of the Sun from the cluster to the rather different Solar orbit would have damaged the outer parts of the Solar system. 
\citet{Pichardo12} carried out their simulations backwards in time
by starting the cluster from the present locus of M\,67. Although spiral arms and the Galactic bar were included in the Galactic potential, the more concentrated 
inhomogeneities in the mass distribution provided by the giant molecular clouds were not represented explicitly. The possibility that the cluster itself had an
earlier, more Solar-like orbit which evolved into its present high inclination orbit via scattering against one or several giant molecular clouds was not suggested.  

Before the analysis of these possibilities of forming clusters in orbits close to the Galactic plane and subsequently scattering them to high altitudes, 
we shall discuss the representativity of the clusters and field star at these heights in some detail in Section 2. In Section 3 orbit calculations for models of the Galaxy with the contributions to the potential from stars and gas, spiral arms, a central bar and GMCs and with detailed consideration of cluster destruction, are introduced and results of these simulations are presented.  The significance of the detailed structure of the GMCs is explored in Section 4. The results will be further discussed in Section 5 where also conclusions are given.
In Appendices, the possible alternative formation of clusters by gas in high-altitude orbits is discussed, and some details of the numerical representation used for the 
gravity potentials are given.

\section{The population of high-altitude clusters}

\subsection{Comparions with stellar distributions}
One realistic explanation for the high-altitude clusters could be that they represent the tail of the z-velocity distribution of a considerable number of open clusters, 
most of which have now been dissolved into the older Thin Galactic disk. The reason why they now stand out would then be a natural selection effect, 
since clusters spending their lives closer to the Galactic plane would dissolve more rapidly. This idea appears commonly in the literature, 
e.g. \citet{Friel95} states: ``The old clusters not only spend their time in the outer disk away from the disruptive effects of giant molecular clouds, 
they spend their time at large distances from the Galactic plane, further enhancing their survivability." 

 A question is then whether one could, 
adopting realistic destruction rates (cf. for example the empirical result of \citet{Wielen71} that about 2\% of all open clusters in the Galactic disk survive beyond 1 Gyr), 
explain the present population of clusters at high altitudes with orbits from a ``normal" velocity distribution of the thin disk population. In the present Section this issue will be discussed from an empirical view
point, on the basis of survey data concerning stars and stellar clusters at different heights above the Plane and their metallicities.

In their UBV star counts towards the North Galactic Pole \citet{Yoshii87} found a scale height of the stellar disk of 250-325 pc. 
The $z$ distribution of the high-altitude clusters in the WEBDA catalogue (\citet{WEBDA}) shows a 
steeper $z$ gradient than the field stars in the old Thin Disk. For the 37 clusters significantly out of the Galactic plane
in that catalogue, i.e. with $|z|$ ranging from 0.2 to 1.0 kpc, we find that the ratio of the number density 
in the $|z|$ interval $400-1000$ pc relative to that in the interval $150-250$ pc,
is smaller by at least a factor of 2 in comparison with the corresponding ratio for field stars.
\citet{Friel95} reviewed the knowledge about old open clusters in the Galaxy and quoted \citet{Janes94} 
who found that the old cluster population is fit by a 375-pc scale-height exponential, 
an appreciably thicker distribution than that of the 55-pc scale-height young cluster population,
but consistent with that found for other old disk populations. 
However, when studying Table 1 of 72 clusters in \citet{Janes94}, we find a clear tendency for the more distant ones
to be at high altitudes, which seems to be due to the difficulty to identify distant open clusters in the Galactic disk, due to extinction and crowding. 
If we confine the sample in \citet{Janes94} to clusters within a Galactic cylinder with a radius of 3 kpc, we find a much steeper $z$ gradient, 
more in agreement with that in the WEBDA catalogue (c.f. also Fig. \ref{clustz}). Although one should not overinterpret the rather inhomogeneous
latter compilation, it seems that the clusters with $|z|>0.2$\,kpc do show a significantly steeper gradient,
i.e. a smaller scale-height than the general stellar field.  Recently, \citet{Buckner15} have traced an increase of the scale height with age of the 
open cluster distribution from the Galactic plane, rising to 550 pc at 3.5 Gyr. They ascribe this tendency to scattering of the clusters from the
Plane in the past.

How does the metallicity z gradient for stars compare with that of open clusters in general? 
\citet{Yoshii87}  traced a stellar metallicity gradient, d[Fe/H]/dz,  of -0.5 kpc$^{-1}$.  The average metallicity for stars of Solar age in the Solar neighbourhood 
is disputed but ranges in the interval -0.2 to 0.0 which then, with the gradient quoted, implies an [Fe/H] at the height of 450 pc of -0.4 to -0.2.
\citet{Cheng12} determined metallicity gradients in the Galaxy as a function of z 
on basis of the Segue survey spectroscopic data. From their Fig. 7 we find a mean [Fe/H] of about $ -0.27$ at z = 450 pc. From the metallicity distributions
of the model of Galactic disk by \citet{Schonrich09}, their Fig. 7, we estimate that about 90\% of the stars  at z = 450 pc are more metal-poor 
than the Sun, a number consistent with the statistics of \citet{Cheng12}. Also, the metallicity
distributions of Schlesinger et al. (2012) from Segue suggest a similar tendency. 
Presumably, these field stars are also older than the Sun at a mean, although, according to \citet{Schonrich09}, the Thin disk still dominates the 
stellar populations at this height. 
Obviously, the stars at heights of 450 pc
above the Galactic plane are at a mean less metal-rich than the metal-rich high-altitude clusters in Table 1.
The open clusters in the plots of \citet{Cheng12} also deviate in that they show systematically higher 
metallicities than indicated by the field-star relations. This departure is also found for the Cepheids. 
In fact, if we adopt the metallicities of open clusters listed in the WEBDA catalogue
we find that the metallicity distributions for clusters in the height intervals $200 < |z| < 500$ pc and $500 < |z| < 1000$ pc depart
significantly from the metallicity distributions at high altitudes for both G and K field stars by \citet{Schlesinger12}, such that the cluster distribution 
has a median [Fe/H] about 0.2 dex higher than the corresponding distributions for the field stars.

In spite of possible selection effects in the data, it seems clear that the metallicity distribution for relatively old open clusters with 400 pc $<|z|<$ 800 pc
is skewed towards high values, as compared with the corresponding distribution for typical field-star metallicities. 
This effect  may well be the result of systematic age differences due to the fact that older clusters (with smaller [Fe/H]) have been dissolved.

The possibilities that the clusters were formed by gas
in high-inclination orbits from the outset seem small, on the basis of statistics of young early-type stars at high altitudes, 
see Appendix A.1, provided that the star-formation-rate in the Disk was not orders of magnitude higher some Gyr ago than 
presently.  In Appendix A.2 we also make some comments on the possible ``unusual" formation scenarios, as reviewed by
\citet{VandePutte10}. 
The high-altitude clusters could also possibly, after formation close to the Galactic plane, have been subsequently scattered
to high orbits by infalling objects like globular clusters, high-velocity clouds, or even supermassive black holes or  
dark-matter sub-halos. Such alternatives are further discussed by Pfister \& Gustafsson (2016, in preparation), who find that
these mechanisms do not seem to be
probable explanations for the Galactic metal-rich high-altitude clusters, but nevertheless
may possibly be important under certain other conditions. 

\subsection{Survival of massive clusters}
Can we estimate the chance for a massive cluster formed in the Galactic disk to survive? We shall here first
attempt an empirical approach towards this problem, assuming that the formation of the M\,67-like clusters just occurred according to the same
principles as other clusters in the Disk. In next Section we shall explore this
question in a more theoretical approach. 

We may estimate the number of
M\,67-like clusters formed in the Galaxy, using the cluster initial mass function according to \citet{Lada03} of 
\begin{eqnarray}
\label{eq4}
dN/dM_{cl} = A\cdot M_{cl}^{-2}.
\end{eqnarray}

\noindent Assuming the total cluster-formation rate $C$ to be independent of time, we find the constant $C\times A$ by integrating Eq.(\ref{eq4}) using a formation rate representative for the Solar neighbourhood (within 600 pc, the distance within which the
surveyed volume is reasonably complete)
of 400 M$_{\odot}$ per Myr for bound clusters with masses M$_{cl}$ in the interval 100 M$_{\odot}$ to $3\cdot 10^4$ M$_{\odot}$ 
from \citet{Lamers06}. This gives $C\times A \simeq 70$ M$_{\odot}/$Myr which leads to the number of clusters formed in the mass interval 
$10^4$ M$_{\odot}$ to $3\cdot 10^4$ M$_{\odot}$ (representing M\,67-like clusters in initial states according to \citet{Hurley05}) of about $5 \cdot 10^{-3}$ per Myr.
 For the period spanned by the clusters in Table \ref{tab:cluster_data} of 5 Gyr, and assuming that all started with masses of at least $1.0\cdot 10^4$ M$_\odot$,
we find that about  1000 clusters of this magnitude should have formed in a Galactic cylinder with a radius of 4 kpc (to match the maximal distance in Table \ref{tab:cluster_data}). 
We find 4 such clusters in the Table. This suggests a fraction of survivals at high latitudes of the total number of clusters produced of about 0.4\%. 
However, since the cluster formation rate, presumably following the star formation rate in general, may have been about a factor of two higher 
4 Gyr ago (guided by the estimates of the history of the star formation rate in the Disk of \citet{Just11} (their model A and Figure 1), this estimate could be decreased to about 0.2\%.

We may alternatively follow a different approach in our estimate:
the known young massive star clusters in the Galaxy with ages less than 20 Myr were listed by \citet{Portegies10}. 
The number of such clusters, defined as having a mass greater than $10^4$ M$_{\odot}$,
is 12, however, as noted by Portegies Zwart et al., these clusters are almost all situated within the Solar quadrant of the Galaxy
which most probably reflects selection effects in the surveys.  Thus, a 
reasonable assumption is that the true number of such clusters in the Galaxy as a whole is at least 40. 
Assuming the mean cluster formation rate some
Gyr ago in the Galaxy to be about two times the present one (again following \citet{Just11}) 
we find that at least about 4000 such massive clusters should have formed per Gyr in the whole Galaxy.
We may estimate from the data in Table 1 that about 12 old metal-rich clusters, still surviving at
high altitudes, were produced per Gyr, This suggests a total fraction of high-altitude survivals of all massive clusters formed
of 0.3\%, which is also consistent with what we obtained from the Cluster Initial Mass Function of \citet{Lada03}, 
Eq. (\ref{eq4})  above. Allowing for the considerable uncertainties involved we 
thus estimate the observed fraction {\bf F}$_{obs}$, of all massive open clusters 
that were formed in the Galactic disk about 4.5 Gyr ago and still survive at heights $|z|>400 $ pc, to be $0.2-0.5\%$.

\section{Scattering in the Galactic disk}

In this Section we shall perform a detailed model study of the possibilities that the high-altitude clusters were formed in low-altitude orbits
close to the Galactic plane, and were later scattered to the high altitudes by secular processes.

\subsection{Encounters with giant molecular clouds and cluster destruction: preliminary considerations}

The effects of encounters with GMCs in the Galactic disk will now be considered. The GMCs have typical masses of
about $5\cdot10^5$ M$_\odot$ (though the masses may extend up to 10 times greater than that) and sizes of typically 40 pc and are located in particular at 
galacto-centric distances from 4 to 9 kpc. Their distribution perpendicular to the Galactic plane
has a $|z|$ scale height of 60 pc - 120 pc with a value increasing 
with the distance from the Galactic centre, the latter value referring
to the Solar circle (data from \citet{Solomon79}, called SSS below, and \citet{Ferriere01}), or even less for the heavier
clouds (\citet{Stark05}). The number of GMCs in the Galaxy was estimated by SSS to be 4000 which with the
mean size given implies a filling factor of the clouds projected onto the Galactic plane of about 3\% between 4 and 9 kpc from the centre 
(for further discussion of these estimates, see Sec. 3.2 below). The GMCs have typical
1D cloud-to-cloud velocity dispersions of about 8 km/s relative to the general Galactic rotation (\citet{Stark89}), probably mainly generated by stellar winds and
supernovae. Locally, however, the scatter may be smaller: \citet{Ramesh94} estimates it to of 3-6 km/s.  
The observed structure of the individual clouds is more reminiscent of sheets or filaments than of spherical blobs (see, e.g. \citet{Allen00} and \citet{Butler15}). 

The number of GMCs in the Galactic disk is high enough for a disk star of about Solar age to come relatively close to a GMC 
several times during its life time. 
If a typical absolute velocity change of a cluster of $v$ would result from each encounter with a GMC, the acquired velocity 
after $n$ encounters might well be $\sqrt{n}v$ if the encounters are statistically independent. The question is then what the value of $v$ might be. 
One way to estimate it is to apply the Rutherford impulse approximation. One may follow \citet{Binney87}, their Eqs. (7.9a) - (7.10b), to show that
the maximum possible addition to the velocity of the open cluster, with a mass significantly smaller that the mass $M$ of the GMC, 
after the scattering in the $z$ direction is
\begin{eqnarray}
v_{z, max}   = V_0 \approx (GM/b)^{1/2} \approx 18 \frac{(M/10^6\,M_\odot)^{1/2}}{(b/10\,{\rm pc})^{1/2}}\, {\rm km/s},  
\label{eq37}
\end{eqnarray}
\noindent where $b$ is the impact parameter and $V_0$ is the relative velocity at infinity. 
Thus, each encounter with a GMC may induce velocity changes of about 10 km/s. However, relative velocities of about 18 km/s 
may not be unrealistic as such, due to the velocity spread of the GMCs and the clusters. Smaller impact parameters
$b$ than 10 pc will not lead to much greater effects since the radius of these relatively diffuse objects is typically 20 pc.  Cloud masses significantly greater
than $10^6$ M$_\odot$ will generate greater speeds, but the mass distribution observed for the GMCs suggests that such
massive clouds are relatively rare (see SSS, and below). 
For vertical motion components of the GMCs, a Disk half-width of 100 pc corresponds to cloud velocities of about 10 km/s. 
In order for collisions with GMCs to be active in bringing a cluster into a high-elevation orbit (i.e. getting a cluster velocity perpendicular 
to the Galactic plane 
of about 30 km/s)
it is obvious either that the cluster or the cloud must depart from the typically low initial relative speeds of the Pop I objects, or that a number of encounters with GMCs must have occurred. These should then also happen to interfere constructively, 
systematically adding to the velocity of the cluster. 

There may be several more complex acceleration mechanisms that hypothetically might affect the speed of the scattered star or cluster. 
One would be due to the transient 
character of the GMCs. If the cluster is falling in towards the centre of a cloud but star-formation occurs in the cloud so that it is dissolved by supernovae explosions before the cluster has passed the cloud centre, 
the absolute momentum excess gained by the cluster during the infall phase may not be fully 
compensated for by retardation during the departure from the centre. It is easy to demonstrate, however, that this mechanism will not lead to
greater contributions to the cluster speed than at most a few km/s as long as the dissolution of the cloud does not take
place very abruptly. 
Another possibility is that the fragmented structure and internal dynamics of the GMC may contribute to the acceleration (or deceleration) of an incoming object. This possibility 
will be further commented on below. Still another mechanism would be that cluster stars are scattered or captured by the encounter with the
GMC but that the core of the cluster picks up momentum and is thrown away at higher speed. 

We must also consider the risk that the tidal forces at a near-by passage by a GMC may 
destroy the cluster. 
A simple measure of the critical distance $b_{crit}$ for a tidal break-up may be
obtained by adopting the impulse approximation and, following \citet{Spitzer58},
estimating the inner kinetic energy change $\Delta(E)$ of the cluster with mass $m$ due to an encounter with a GMC:

\begin{eqnarray}
\Delta(E) =  4/3 \cdot G^2  m M^2 r_s^2/(b^4 V_0^2),
\label{eqnspitzer}
\end{eqnarray}

\noindent where $b$ is the impact parameter (at infinity), $G$ Newton's constant of gravity, $V_0$ the relative velocity of the GMC and the cluster (at infinity), 
and $r_s$ is the root mean-square radius of the cluster. We assume that the cluster is in virial equilibrium 
such that the inner kinetic energy $E$ of the cluster is half of the absolute gravitational one, i.e.

\begin{eqnarray}
E = \gamma m^2 G /(2  r_s).
\label{eqnvirial}
\end{eqnarray}

\noindent The numerical coefficient 
$\gamma$ depends on the distribution of stars in the cluster. Following \citet{Spitzer58}
we set $\gamma = 0.5$ and obtain

\begin{eqnarray}
\frac{\Delta(E)}{E} = \frac{16 GM^2r_s^3}{3m b^4 V_0^2}.
\label{eqnequilib}
\end{eqnarray}

\noindent We now assume that the cluster breaks up if $\Delta(E)/E > 1$ and thus obtain for the critical impact parameter $b_{crit}$ for which break-up 
is expected to occur·

\begin{eqnarray}
b_{crit} = 16 {\rm pc} \frac{(M/10^6\,{\rm M}_\odot)^{1/2}}{(m/10^4\,{\rm M}_\odot)} \cdot \frac{(r_s/{\rm \,1\,pc})^{3/4}}{(V_0/{\rm 10\,km/s})^{1/2}}.
\label{bcond}
\end{eqnarray}

\noindent More detailed numerical simulations of cluster destruction are presented in Sec 5.2.

\subsection{Scattering of M\,67 to its high altitudes: a global synthetic approach.  The detailed Galatic model}

As discussed above, the spread of open clusters vertically relative to the Galactic plane may be the result of 
a process where many GMC collisions are involved. Also other phenomena may play important roles such as gravitational
perturbations by
spiral arms in the Galactic disk or the Bar. In order to obtain realistic numbers on the probability of the scattering of stars and clusters to high latitudes in this ``Galactic landscape" we have carried out numerical simulations of orbits
of individual test particles moving in a model galaxy with stars, spiral arms, a central bar and GMCs included.
Each of the test particles represents a star or a cluster. The destruction of the clusters by tidal interaction is considered 
schematically but is also studied in some detail (see Sections 3.3 and 5.1, below).  

The test particles move in an axially symmetric Galactic potential according to Potential I of \citet{Binney12}, adjusted to be consistent with a circular speed of 220 km/s
at $R_0 = 8$ kpc and with some corrections for added masses 
described below. The mass distribution has components from
Thin and Thick stellar disks, a gas disk, and a stellar and dark spheroid representing the Bulge and the Halo. (We have also made experiments with modifications of the z-gradient
of the potential by $\pm10\%$ to explore the effects of its uncertainties.) To this we have added two {\it stellar} spiral arms, 
following the recipe of Pichardo et al. (2003, 2012), with a pitch angle of 15.5 degrees, a radial scale length exponential mass decrease along the arms of 3.9 kpc,
a mass of $4\cdot 10^9$ M$_{\odot}$ and a constant pattern speed of 24 km/s kpc$^{-1}$. Each arm is represented numerically by 100 oblate inhomogeneous spheroids 
with semi-major and minor axes of 1000 pc and 500 pc respectively, and with a mutual distance between the spheroid centres of 500 pc.  
The bar is represented by a prolate inhomogeneous spheroid
following \citet{Pichardo03} and \citet{Pichardo12}, with a total mass of $1.6\cdot 10^{10}$ M$_{\odot}$, with density scale lengths in the Galactic disk of 1.7 kpc 
(along the major axis, which corresponds to an effective boundary of the bar at 3.13 kpc) and  
0.54 kpc, respectively, with the axis perpendicularly to the plane also assumed to be 0.54 kpc, and with an angular speed of 55 km/s, kpc$^{-1}$. For the spheroids, representing the
Bulge as well as for those of the Spiral arms, we adopted a linear density variation with radius, just as that used by Pichardo et al. For motivations and uncertainties
in the parameters of these representations, see Pichardo et al. (2012). Both the set of spiral arms and the bar are assumed to be stationary rotating structures with constant angular speeds.  
It should be noted that the different elements in the model, including the overall Galactic potential, the spiral arms and the Bar, are not dynamically consistent from the outset, 
nor are they allowed to relax to a dynamically consistent configuration. This kinematical, sooner than dynamical, model has, however, the virtue that it may be suppposed to describe a 
reasonably realistic semi-empirical potential. We consider it probable that its lack of consistency does not lead to extra scattering of the test particles. Sooner could such scatter be
introduced if the model was made more dynamically consistent, before the various elements had relaxed.  
The numerical representation of the gravitational forces from these different components is described in Appendix B.

The GMCs were initiated from randomly chosen points in the spiral arms, within a distance of $\pm 50$ pc from the median line of the arm, and with a number 
density decreasing exponentially along the arm to match that of the stellar arm itself. As an alternative, the GMCs were generated randomly in the disk, also outside the 
arms, however, again with an exponential decrease of the formation probability 
with the distance from the galactic centre. It was found that the effects on the final statistical properties of the stellar/cluster
orbits in the simulations of shifting between these two alternatives were astonishingly small -- less than 2 km/s in the final velocity scatters for stars at a distance of about 8 kpc from the Galactic Centre. 
Subsequently, only results for the first alternative, e.g. GMCs originating close to the spiral arms, are given. The GMCs were given initial
velocities with gaussian distributions and according to a characteristic velocity ellipsoid with axes of 7 km/s,  in agreement with observations of \citet{Stark84}, see also \citet{Larson79} 
and \citet{Fukunaga84} but greater
than those of \citet{Ramesh94}; however, experiments with values lower than 7 km/s did not lead to very significant changes of our final results. 
We found that with this initial velocity distribution
our ensemble of model clouds showed a z distribution with e-fold
decrease relative to the plane at a $z$ of $\pm 75$ pc which is close to the observed value of SSS. Again, however, the final result is relatively independent of our assumed
starting-velocity distribution for GMCs. 
The maximum mass of cloud $\# i$, $M_i$, was selected randomly with a distribution function (see \citet{Williams97}, and \citet{Hopkins12}, HQM below) of
\begin{eqnarray}
N(M_i) dM_i = {\rm const.}\times M_i^{-1.8} dM_i, \,\, 5 \leq {\rm log}(M_i/M_\odot) \leq 7.0,
\label{eq371}
\end{eqnarray}
while no clouds were generated with maximum mass outside this interval. In our standard models the GMCs were just added to the homogeneous disk
with the total disk mass (and thus gas density) correspondingly globally reduced. In alternative models, mass conservation was considered more locally,
see below. The evolution of the clouds was considered as follows:
Each GMC was given an individual 
life time corresponding roughly to a few free-fall times of the cloud, i. e. totally 40 million years, with a mass increasing to a value $M_i$ in 20 million years, and then decreasing 
to zero in another 20 million years (see HQM, in fair agreement with the simulations of \citet{Krumholz06} and \citet{Goldbaum11}, following a parabolic mass evolution:
\begin{eqnarray}
M_t = \left\{-0.25 \cdot [(t-t_0)/10^7]^2 + (t-t_0)/10^7)\right\}\cdot M_i,
\label{eq379}
\end{eqnarray}
where $t_0$ is the time of the formation of the GMC and $t$ the running time, both in years. The evolution effects
made the distribution of average masses vary between the limits $2/3\cdot 10^5 -2/3\cdot 10^7$ M$_{\odot}$ while the distribution for the 
clouds in general has the limits $0 - 10^7$ M$_{\odot}$. We note, however, that \citet{Fukui10} found a maximum GMC mass of 
$5\times 10^6$ M$_{\odot}$ in nearby galaxies. The mass distributions of Williams \& McKee and HQM continue below our
lower mass limit by one order of magnitude; these numerous lighter clouds were not included explicitly in the calculations of orbits,
which in general are less strongly affected by these clouds, but included in the contribution from the gas disk to the general Galactic potential.  
The values of $t_0$ for the individual clouds
were taken at random through the time interval from 0 to 4.6 Gyr, adopting a probability distribution constant in time, i.e. we assume the number of 
GMCs in the Galaxy not to vary systematically with time.  In the standard case, each GMC was represented by a Plummer sphere (\citet{Plummer11}) 
with a typical radius $R_c$ of about 20 pc (SSS), and $R_c$ scaling with the square root of the cloud mass as suggested by HQM, i.e.
\begin{eqnarray}
R_c(M_t) = 20 (M_t/5\cdot 10^5\,M_\odot)^{1/2} \rm{pc}.
\label{eq3711}
\end{eqnarray}

\noindent In some orbit simulations we also kept one tenth of the GMC mass concentrated into a homogeneous core with a radius of 1 pc to take the existence of 
condensed cores in the centres of the GMCs into reasonable consideration (see, e.g. \citet{Bergin96}). The modifications of
the final results were, however, marginal. The effects of other modifications of the internal structure of the GMCs and their surroundings were, however, found to be considerable.
These will be discussed in Section 4. 

The total number (4000) of GMCs in the present Galactic disk as estimated by SSS
mainly includes clouds with masses  $\ge 10^5$ M$_{\odot}$. 
\citet{Williams97} in their Table 4 favour a higher number
of clouds, but most of these clouds have lower masses than the lower limit of our mass interval (and that of SSS),
while if we limit the interval to our effective range of average masses from $6.7\cdot 10^4 -6.7\cdot 10^6$ M$_{\odot}$ 
we find numbers of 2000-3000 from the distributions of Williams \& McKee. 
We note, however, that a very considerable fraction of the clouds included in the study by Williams \& McKee were still undetected, 
and only statistically and schematically corrected for. Willams \& McKee have adopted 
the value of $1.0\cdot 10^9$ M$_{\odot}$ for the total Galactic molecular mass. 73$\%$ of this mass is then found in clouds with masses 
above $6.7\cdot 10^4$ M$_\odot$. 
The corresponding total GMC mass given by SSS is $2\cdot10^9$ M$_{\odot}$, while \citet{Nakanishi16} found an H$_2$ mass of $8.5\cdot 10^8$ M$_{\odot}$. We here adopt the value
of Williams \& McKee for the total mass and then find, when reducing it to represent our mass interval,  a total number of 
clouds of 2500 at present in the Galaxy. 
With a cloud life-time of 40 Myr and assuming the present density of GMCs to be representative for the last 4.6 Gyr in the Galaxy,
our value for the total mass of the molecular gas corresponds to 
an ensemble of altogether about 300,000 GMCs, the action of which was included in our simulations.  
As an alternative, we also explored the results of increasing this number to 460,000, which then corresponds to the figures given
by SSS.

In our rather complex Galaxy potential the molecular gas model clouds were moving, 
however the cloud-cloud interaction was not included in calculating the cloud orbits.
As for the GMCs the test particles representing clusters or stars were expelled from the Galactic plane and their initial radial distances (between 4 and 9 kpc) 
from the Galactic centre and velocities
were chosen randomly. Our main ambition has been to study the velocity distribution and the z-distribution for stars and clusters around the Solar circle, while most
stars form inside that due to the exponential density profile of the gas disk. Therefore, to get an optimal statistics we have biassed the distribution of test particles by 
giving an equal probability
for their origins for every given value of their galactocentric distances $R_0$, 4 kpc $<\,R_0\,< $ 10 kpc, and next, in the final calculation of distributions, means and 
standard deviations, given the particles different weights $w_p$ according to $w_p \thicksim$ exp($-R_0/4800$) pc, following the exponential gas disk of 
Model 1 of \citet{Binney12}. It should be noted that the starting points for the particles were not correlated with the GMCs, except for the general concentration towards the Galactic disk. This statistical independence may underestimate the effects of GMCs
on clusters and stars which are generally formed in dense gas clouds.

The particles were given random initial velocities according to a spherical velocity ellipsoid relative to the local circular rotation speed in the Galactic potential 
with a gaussian spread of 7 km/s in the three velocity components U, V and W. The orbits of the test particles were followed 
for $4.6\cdot 10^9$ years. The number of test particles $N_*$ was in most runs typically chosen to be 500 - 1000 in order to 
secure enough of orbits for reliable statistics, e.g. on the resulting velocity distributions and the distribution of distances from the Galactic disk
at the end of the integration. In some runs, $N_*$ was lowered to 200. 

The orbits were obtained from the equations of motion (for a test particle of unit mass) in cylindrical coordinates:
\begin{eqnarray}
\ddot{R}-R\dot{\phi}^2 = - \frac{\partial \Phi}{\partial R} + F_R
\\
\ \frac{d}{dt} (R^2\cdot\dot{\phi}) = F_\perp \cdot R
\\
\ \ddot{z} = - \frac{\partial \Phi}{\partial z} + F_z.
\label{eqn3710}
\end{eqnarray}

\noindent Here, $R$ is radial coordinate of the particle as measured from the Galactic centre in the Galactic plane, $\phi$ is the corresponding angular coordinate,
measured relative to a fixed direction in space, $\Phi$ is the smooth axisymmetric gravitational potential, $F_R$ and $F_\perp$ are the components of the sum of the forces, not represented by the potential $Phi$, that affect the motion of the particle. These components are directed away from the Centre and  perpendicularly to that direction, respectively, 
and are parallel to the Galactic plane. $F_z$ is the corresponding force component in the $z$ direction. 

Before proceeding to calculating the orbit numerically, we integrate the second of the equations above to obtain
\begin{eqnarray}
R^2\cdot\dot{\phi} = \int_0^t \! F_\perp \cdot R \, \mathrm{d}t + const.
\label{eqn3711}
\end{eqnarray}
The integration constant is determined by the initial conditions. Next, in order to
guarantee a full angular-momentum versus torque balance, 
$\dot{\phi}$ as obtained from Eq.(\ref{eqn3711}) is substituted into
Eq.\,(10), leading to the system
\begin{eqnarray}
\ddot{R}=\left\{\int_0^t \! F_\perp \cdot R \, \mathrm{d}t + const.\right\}^2R^{-3} - \frac{\partial \Phi}{\partial R} + F_R
\\
\ \frac{d}{dt} (R^2\cdot\dot{\phi}) = F_\perp \cdot R
\\
\ \ddot{z} = - \frac{\partial \Phi}{\partial z} + F_z.
\label{eqn3712}
\end{eqnarray}

\noindent In practice, the system was solved for the variable $r=R-R_0$ where $R_0$ was chosen to be the initial $R$ coordinate of the test particle
 and the variable $\varphi =\phi -  R_0\cdot\omega_0$ where $\omega_0$ is the angular speed of 
rotation at $R_0$.
During the integration, also velocities $U$, $V$ and $W$ were calculated.
For this system, stable solutions were obtained for integration times extending to at least $5\cdot10^9$ years by standard MATLAB routines 
such as {\it ode23s} based on a second-order Rosenbrock formula (\citet{Shampine97}), as
was demonstrated by comparison to detailed integration using the 15th order RADAU integrator by \citet{Everhart85}).
The particle was deleted from the {\it cluster} statistics if it ever came so close to a
GMC, and with such small relative velocity that the condition Eq.(\ref{bcond}) for disruption of the cluster was fulfilled. In adopting this
criterion, however, we used $V_n$, the relative velocity when the particle was at its minimum distance from the GMC, instead of $V_0$, 
the relative velocity at infinity. It is easy to prove that applying Eq. (\ref{bcond}) in this way to calculate a critical distance $b_n$ and deleting all clusters within
distances $b_n$ from the GMCs will lead to a systemactic overestimate of the destruction rate of the clusters: all
clusters with $b < b_{crit}$ will (in the two-body case) come closer to the corresponding GMC than $b_n$. The overestimate of the destruction rate
will mainly be significant of low-velocity encounters. 

For test particles that fulfilled this criterion of cluster destruction a flag was set but 
we continued the orbit calculation for totally 4.6 Gyr, in order to be able to apply the
results also in comparsions with observations for individual field stars. 

We performed test simulations with and without the effects
of GMCs, spiral arms and the Galactic bar. As a standard, 500 test particles were
followed for 4.6 Gyr in the model system in every run. The calculations were performed with the Tintin 2560 core cluster and the Milou 3338 core cluster at 
Uppsala Multidisciplinary Centre for Advanced Computational
Science (Uppmax). In order to get satisfactory statistics for the calculation of the number of test particles ending up at high latitudes, we performed
several of the runs repeatedly. The calculations were run in parallel in a simple manner 
such that the orbit for each test particle was run on its own core. Typical runs with $N(GMC)=300,000$ took about 40 hours on each core.

\subsection{The cluster destruction in detail}

We have tested the adequacy of Eq.~(\ref{bcond}), as a basis for estimating the risk of cluster destruction, by
numerical $N$-body simulations, using the gravitational $N$-body code {\tt NBODY6}, \citet{Aarseth03}.
In order to make the problem of modelling the encounters tractable, we consider
only three initial cluster models.  Inspired by the models of \citet{Hurley05}
we started each of these models with an initial mass of
$2.6\times 10^4\,M_\odot$, made up of 36\,000 stars distributed in mass
according to a \citet{Kroupa93} IMF.  
The stars were spatially distributed according to a \citet{Plummer11} distribution.
The choice of initial half-mass radius is somewhat arbitrary.  \citet{Hurley05} choose an initial half-mass radius of $r_{\rm h,i}\simeq 4\,{\rm pc}$,
which with the tidal field that they adopt ensures that the cluster is close to
filling its tidal radius at formation time.  However they point out that an
adequate model of M\,67 can be made using a smaller initial half-mass radius of,
say, $r_{\rm h,i}\simeq 1\,{\rm pc}$, as in 
\citet{Hurley01}.  In this case the cluster initially
does not fill its tidal radius but evolves with a shorter dynamical timescale
such that by a time of $4\,{\rm Gyr}$ both models have rather similar
structures.  We therefore considered models with initial half-mass radii of
1\,pc, 2\,pc and 4\,pc.  This covers the range of initial conditions indicated
by Hurley et al.  The significance of the choice of initial half-mass radius is
discussed further at the end of this Section. 

The initial velocities were chosen so as to give a virial ratio of 0.5 in isolation.  We
chose to treat the stars in our simulations as point particles of constant mass
and to ignore primordial binaries.  The reason for this was two-fold.  Firstly,
the resulting simulations are computationally considerably more
straightforward, both in terms of total runtime and reliability.  Secondly, and
more significantly, both the inclusion of primordial binaries and stellar
evolution increase the variability of a cluster's evolution; that is, they make
the behaviour more stochastic.  We were interested in isolating the dynamical
effects of the encounters with GMCs, hence we chose to make our simulations
as simple as possible.

We tested both the immediate effects of encounters with GMCs on stellar clusters
and the subsequent evolution of the post-encounter clusters using a two-step
process.  In the first step, the encounter of a cluster with a GMC was
modeled.  The GMC was treated as a Plummer potential of total mass $M$,
interacting with the cluster only through the gravitational force that it exerts
on the stars; the force of the stars on the cloud was  neglected.  The cloud was
set up at an initial spatial position of $(x,y,z)=(-100\,{\rm pc},-b,0)$ relative to the cluster, which was
initially at rest at the origin.  In addition, the cloud was given an initial
velocity with respect to the cluster of $(V_x,V_y,V_z)=(V_0,0,0)$.  The cluster was
evolved in this varying potential until the cloud had travelled $200\,{\rm pc}$
from its starting position.  A grid of values of GMC mass $M$, impact parameter
$b$ and relative velocity $V_0$ were modeled, as summarised in
Table~\ref{tab:destructionModelGrid}.  This part of the simulation was carried
out with no external force; i.e. neglecting any effects of the Galactic tidal
field.

\begin{table}
\caption{Parameters of the cluster destruction simulations.}
\begin{tabular}{lllll}
\hline
Parameter                                        & \multicolumn{3}{l}{Values}            \\
\hline
Cluster half-mass radius $r_{\rm h}/{\rm pc}$    & 1&2&4 \\
GMC mass $M/10^6\,M_\odot$                       & 0.1& 0.5& 1& 5\\
Impact parameter $b/{\rm pc}$                    & 10 & 20 & 40\\
Velocity at infinity $V_0/{\rm km\,s^{-1}}$ & 10 & 20 \\
\hline
\end{tabular}
\label{tab:destructionModelGrid}
\end{table}

Once the cloud had reached a position of $(100\,{\rm pc},-b,0)$ the simulation
was stopped.  The cloud potential was removed and the simulation re-centered on
the centre of momentum frame of the stars.  A linearised Solar circle Galactic
tidal field was imposed following the method described in \citet{Aarseth03}.
and the simulations continued until the cluster had evaporated.

\begin{figure}
\resizebox{\hsize}{!}{\includegraphics{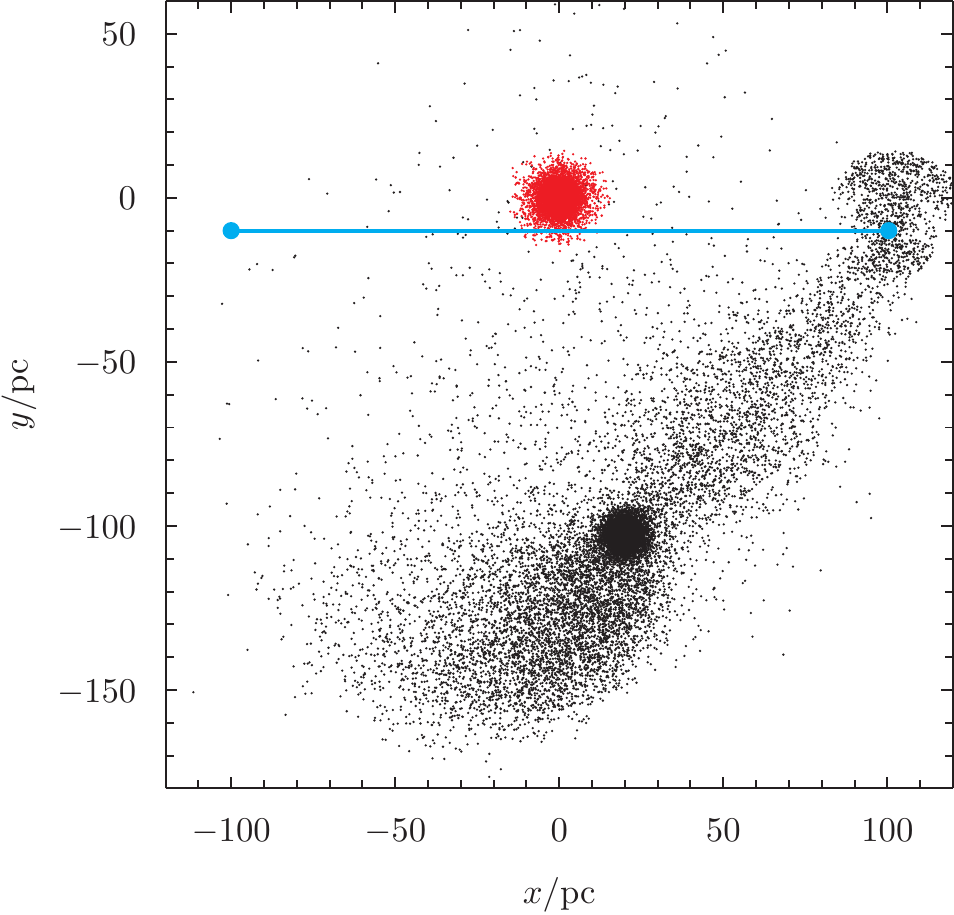}}
\caption{The postions of stars projected onto the cloud-cluster orbit plane in one
of our cluster-destruction simulations.  Red/gray dots show the initial positions
of stars in our model cluster.  The black dots show the positions of stars
at a time of 20\,Myr, just before the GMC is removed from this simulation; i.e at the end of step
one of the two-step process described in Section 5.2.  The positions are
plotted in the frame in which the cluster centre of mass is initially at
rest.  The horizontal line shows the motion of the GMC in this frame from left to right, with the
end points showing its position at the times when it is inserted
and removed, and the encircling dashed line its half-mass radius.  The star cluster in this example simulation had an initial half-mass radius $r_{\rm
h,i}=2\,{\rm pc}$ and interacted with a cloud of mass $M = 3\times
10^5\,M_\odot$ and half-mass radius of 5 pc, moving at a relative velocity at infinity of $V_0 =
10\,{\rm km\,s^{-1}}$. The impact parameter was $b = 10\,pc$. 
}
\label{snapshot}
\end{figure}

The outcome of the first step of one of the cluster destruction
simulations can be seen in Fig. \ref{snapshot}.  It can be seen that the cluster has
been accelerated in the $-y$ direction.  Two tidal tails of stripped stars
are visible, as well as a small group of stars which have become entrained
in the GMC and are visible as the small black halo towards the top right of
the figure.  The majority of stars in the tidal tails and the small halo
have become unbound from the cluster, and hence are removed by the tidal
field of the Galaxy when it is imposed.  However the majority of the
cluster has remained bound and can be identified as the solid black object
located at approximately $y=-100\,{\rm pc}$.  This model cluster survives the
encounter with the loss of aproximately 8000 stars from its original 
36\,000.

\subsubsection{Results of $N$-body simulations of cluster/GMC encounters}

\begin{figure}
\resizebox{\hsize}{!}{\includegraphics{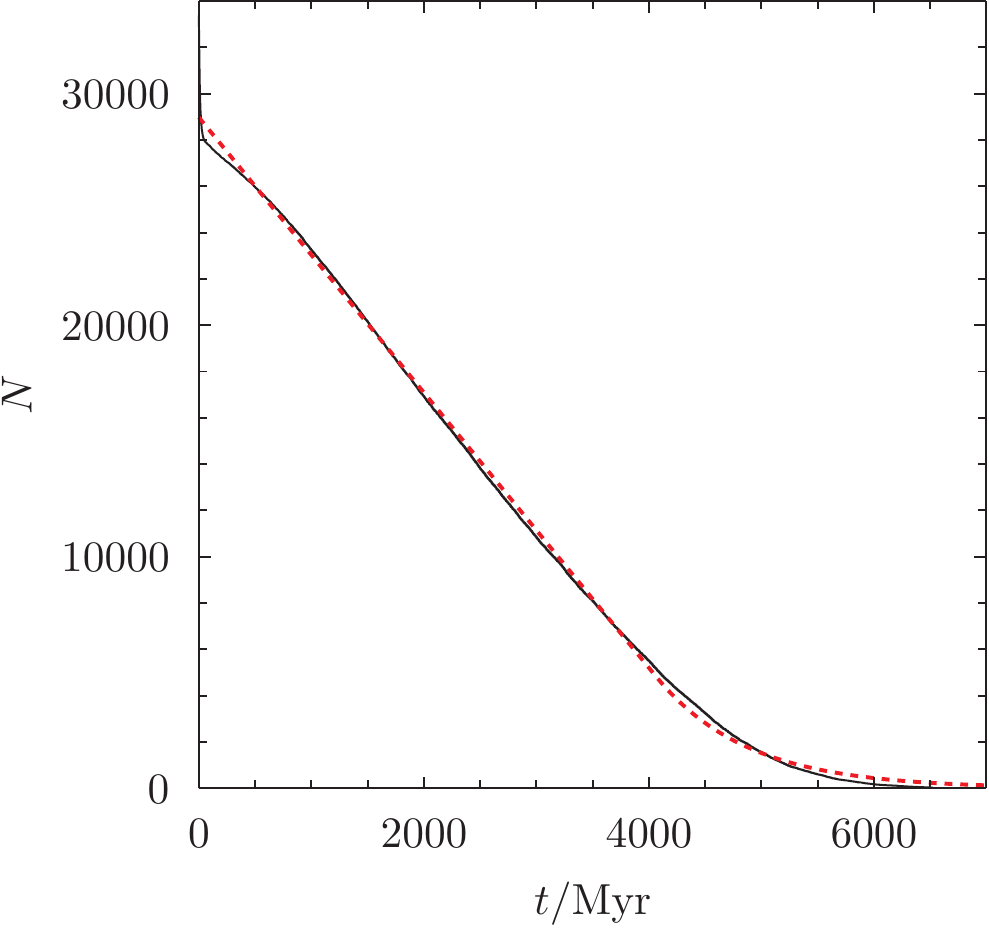}}
\caption{The evolution of the number of stars $N$ with time $t$ in one of our
cluster destruction simulations.  The example simulation had an initial
half-mass radius $r_{{\rm h},i}=2\,{\rm pc}$ and interacted with a cloud of mass
$M=3\times 10^5\,M_\odot$, moving at a relative velocity at infinity of
$v_\infty=10\,{\rm km\,s^{-1}}$.  The impact parameter was $b=10\,{\rm pc}$.
The thin, black, solid line shows the evolution of $N$ from the $N$-body
simulation.  The red, dashed, thick line shows the best-fit fitting function.}
\label{fig:fitN}
\end{figure}

In a small number of cases the cluster was completely disrupted by the
encounter; that is, there was no discernible bound object remaining.  In all
cases such clusters were those predicted to be disrupted by the criterion of
Eq. (\ref{bcond}).  All other cluster models were evolved, losing stars in the
Galactic tidal field, until only a few tens of stars were left.  We found that
the evolution of the total number of stars $N_i$ in cluster $i$ as a function of
time $t$ was in each case fairly well fit by a function of the form

\begin{equation}
N_i(t)=\left\{
\begin{array}{lc}
      N_{0,i} - m_i t     & t<t_{{\rm b},i} \\
      (N_{0,i} - m_i t_{{\rm b},i})\exp\frac{-(t-t_{{\rm b},i})}{N_{0,i}/m_i+t_{{\rm b},i}} & t>t_{{\rm b},i} \\
\end{array}
\right.,
\label{eqn:fitN}
\end{equation}
where the initial number of stars, $N_{0,i}$, the initial slope, $m_i$, and the
break time at which the function transitions from a straight line to an
exponential decay, $t_{{\rm b},i}$ are parameters of the fit, which we obtained
using a least-squares fitting procedure.  Fig.~\ref{fig:fitN} shows an example
fit.

\begin{figure}
\resizebox{\hsize}{!} {\includegraphics{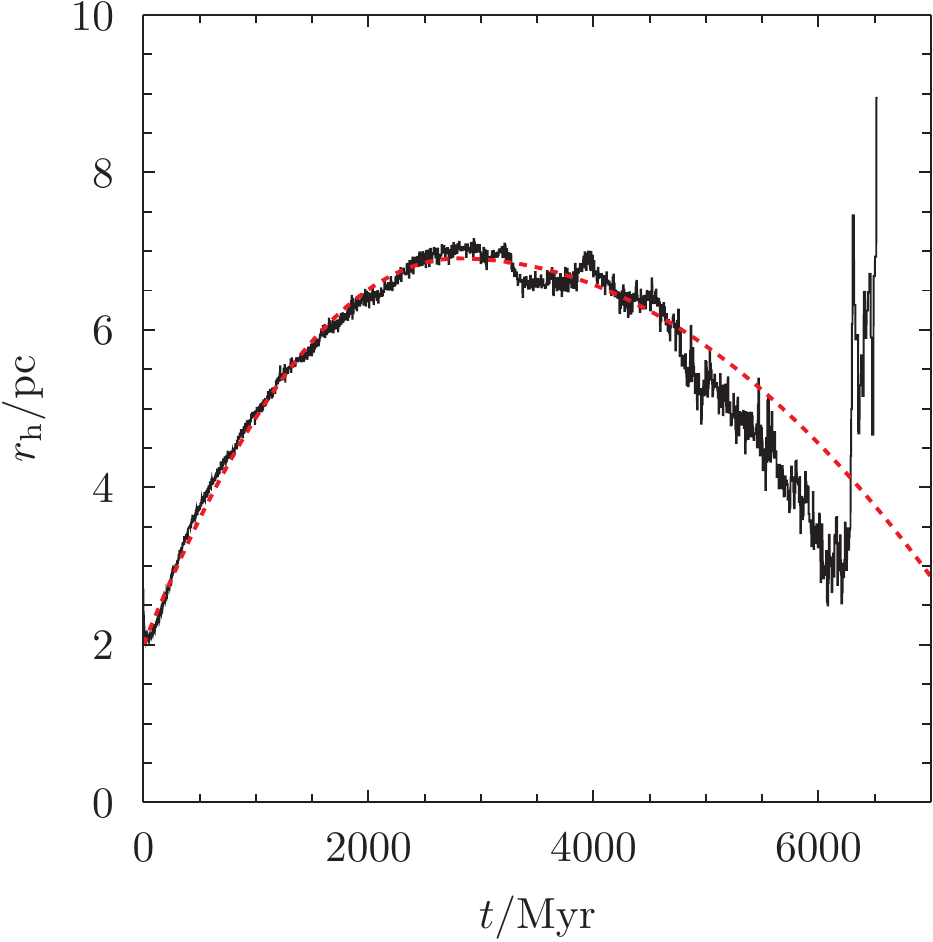}}
\caption{The evolution of the cluster half-mass radius $r_{\rm h}$ with time $t$
in one of our cluster destruction simulations.  As in Fig.~\ref{fig:fitN},
the example simulation had an initial half-mass radius $r_{{\rm h},i}=2\,{\rm
pc}$ and interacted with a cloud of mass $M=3\times 10^5\,M_\odot$, moving at a
relative velocity at infinity of $v_\infty=10\,{\rm km\,s^{-1}}$.  The impact
parameter was $b=10\,{\rm pc}$.  The black solid line shows the evolution
of $r_{\rm h}$ from the $N$-body simulation.  The red dashed line shows
the best-fit fitting function.}
\label{fig:fitRh}
\end{figure}

The evolution of the half-mass radius with time is slightly more complex than
that of the total number of stars since it initially increases owing to internal
dynamical processes, then decreases once the cluster fills its tidal radius and
starts to lose mass.  After some experimentation we found that we could obtain
an adequate fit from a broken quadratic function, although the fit is in general
less accurate than that to the total number of stars, particularly in the later
parts of the evolution.  The fitting function that we adopted is

\begin{equation}
r_{{\rm h},i}(t) = \left\{
\begin{array}{lc}
   a_{1,i}t^2-2a_{1,i}t_{{\rm to},i}t+c_{i}                               & t<t_{{\rm to},i}\\
   a_{2,i}t^2-2a_{2,i}t_{{\rm to},i}t+(a_{2,i}-a_{1,i})t_{{\rm to},i}^2 + c_{i}  & t>t_{{\rm to},i}\\
\end{array}
\right .
\label{eqn:fitrh}
\end{equation}
where the fit parameters are the two curvatures, $a_{1,i}$ and $a_{2,i}$, the
initial value, $c_i$, and the turnover time, $t_{{\rm to},i}$.  An example fit
is shown in Fig.~\ref{fig:fitRh}.

Having obtained fitting parameters for each of our clusters, we investigated
their behaviour as a function of the analytically predicted fractional change
in cluster binding energy. 
We have here used a slightly different definition of the predicted fractional change
in binding energy from that given in Eq. (\ref{eqnequilib}):

\begin{equation}
\delta_E = \frac{8GM^2r_{\rm h}^3}{3mb^4V^2}
\label{eqdelta}
\end{equation}

\begin{figure}
\resizebox{\hsize}{!}{\includegraphics{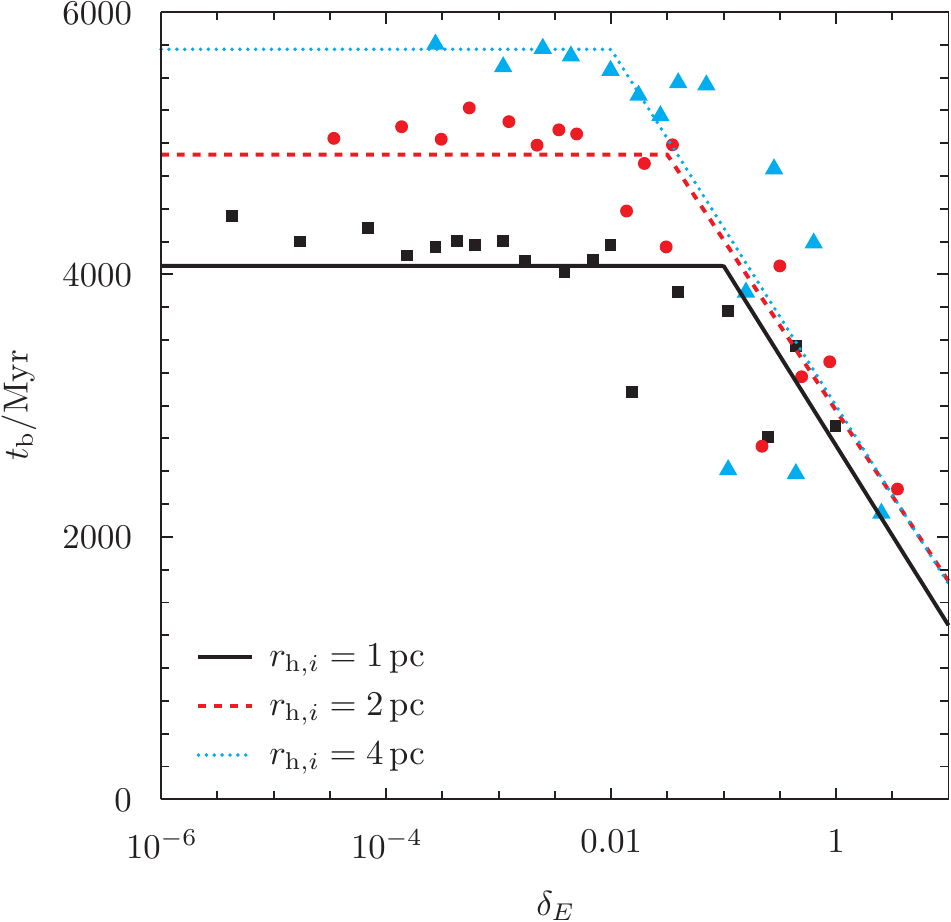}}
\caption{The behaviour of $t_b$, the break time in the fits of $N$ with $t$, as
a function of predicted $\delta_E$.  Clusters with initial half-mass radii of
$1$, $2$, and $4\,{\rm pc}$ are shown by black squares, red circles and blue
triangles respectively.  The solid black, dashed red and dotted blue lines show
the fits adopted to the results for half-mass radii of $1$, $2$, and $4\,{\rm
pc}$ respectively.}
\label{fig:fitFit}
\end{figure}

\noindent Note that we here use the
cluster half-mass radius $r_{\rm h}$, directly provided by our $N$-body simulations, 
rather than the root mean-square radius $r_{\rm s}$. 
The fractional change in the binding energy is a quantity 
which relates to the degree of cluster dissolution but its exact absolute value is
not of importance here. Therefore, the scaling difference between $\Delta E/E$ and $\delta_E$
is not significant. 
The reason for using $\delta_E$ rather than $\Delta E/E$, as measured from the simulations, 
is that we wish to adapt our results to
encounters which we have not made $N$-body models of.  $\delta_E$ can be calculated for any
encounter, whereas $\Delta E/E$ must be obtained from $N$-body simulations. 
Having plotted all the fitting parameters as functions of
predicted $\delta_E$, we found that there was a consistent trend.  For each
value of $r_{{\rm h},i}$, up to a given value of $\delta_E$, the values were
constant, up to some scatter.  This implies that weak encounters have very
little effect on the evolution of the cluster.  At larger values of $\delta_E$
we found that the fitting parameters varied, up to some much larger scatter,
approximately linearly with $\log_{10}\delta_E$.  Hence we fit, for each of
our original fitting parameters, constant values breaking to a straight line as
a function of  $\log_{10}\delta_E$; see Fig.~\ref{fig:fitFit} for an
example.  This process allows us to, for any encounter, having calculated
$\delta_E$, generate fit parameters for the evolution of $N$ and $r_{\rm h}$
with time and hence easily simulate the evolution of the cluster.

\subsubsection{Synthetic cluster encounters}

In order to use our formalism to predict the effect of encounters on our
clusters, we make two basic assumptions.  The first is that we can treat
multiple encounters by summing the fractional change in cluster binding energy
$\delta_E$.  The second is that, following an encounter at time $t$, the
properties of the cluster are the same as those of the same cluster with the
same $\delta_E$ had the encounter occurred at time $t$.  Having done this,
we follow the history of each cluster in our global synthetic simulation.
Starting at time zero, the cluster is taken to have $\delta_E=0$.  When
it first encounters a GMC we use the velocity and distance at closest approach
to generate $b$ and $V_0$ for the encounter, and our fits to an unperturbed
cluster evolution to obtain $m$ and $r_{\rm h}$.  We then use
Eq.~(\ref{eqdelta}) to calculate the change in binding energy.  Finally we
step forwards to the next encounter and repeat the process, but this time using
the fits for the newly increased value of  $\delta_E=0$ to obtain the cluster
properties at the time of encounter.


\subsection{Resulting stellar orbits and velocity dispersions in the Galaxy simulations}

\begin{figure}[htbp]
\centering
\resizebox{\hsize}{!} {\includegraphics{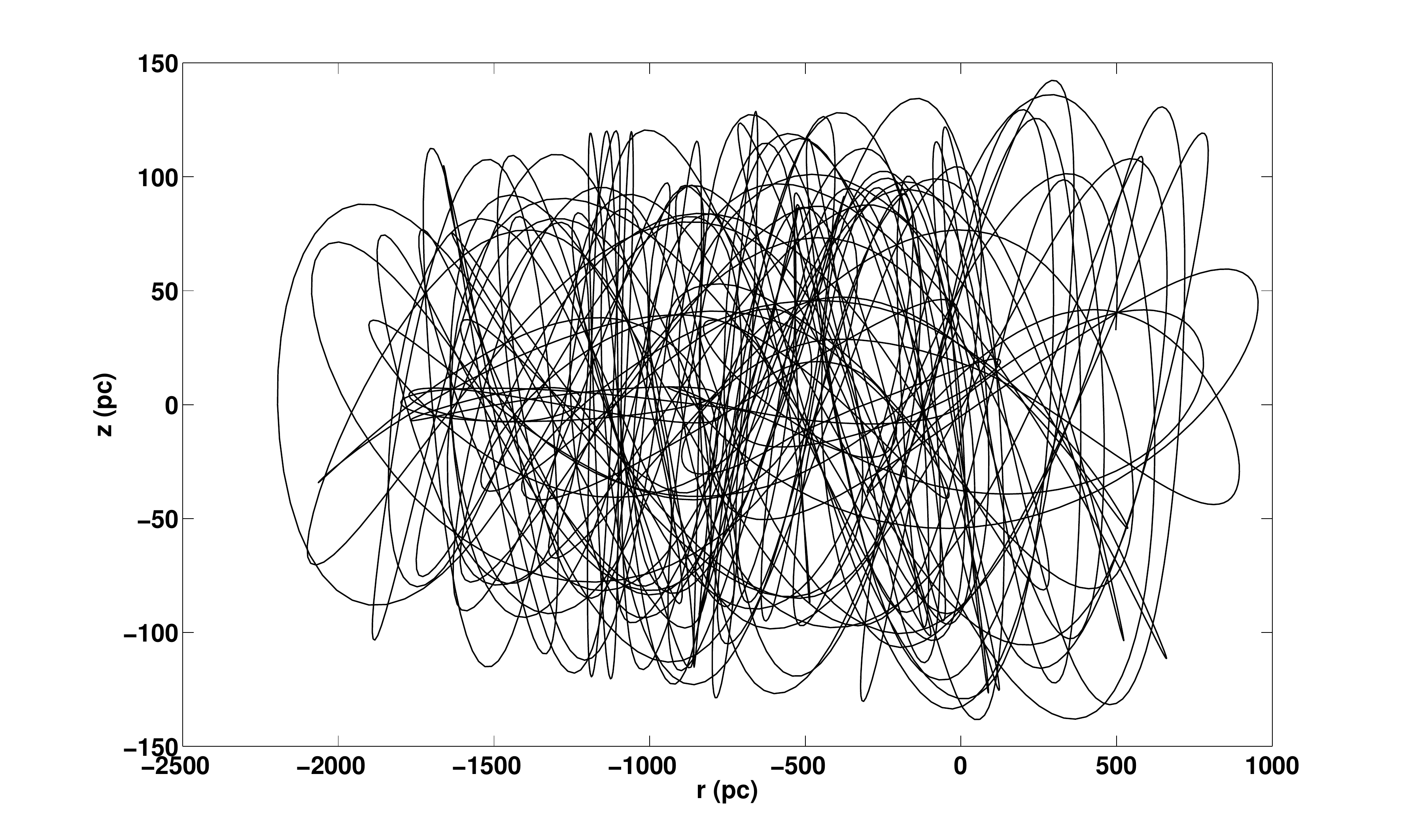}}
\caption{An orbit of a test particle from a simulation with Bar, GMCs and Spiral arms (BGS) in the $R - z$ plane, followed for 4.6 Gyr. } 
\label{orbit1}
\end{figure}

\begin{figure}[htbp]
\centering
\resizebox{\hsize}{!}{\includegraphics{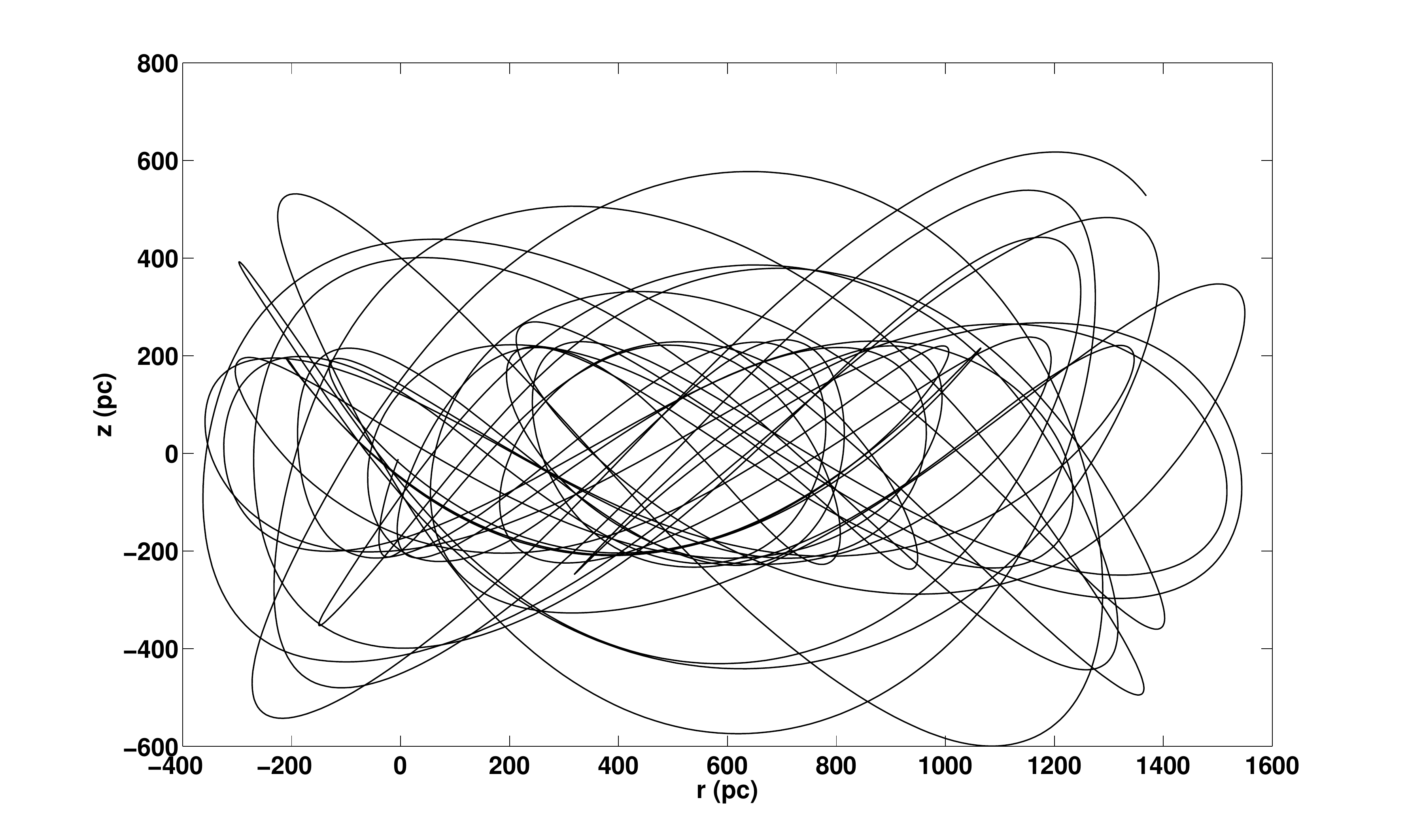}}
\caption{An orbit of a test particle from a BGS simulation in the $R - z$ plane, followed for 4.6 Gyr. This particle is one of the few that reached a height above 
the Galactic disk above 400 pc.  } 
\label{orbit7}
\end{figure}

\begin{figure}[htbp]
\centering
\resizebox{\hsize}{!}
   {\includegraphics{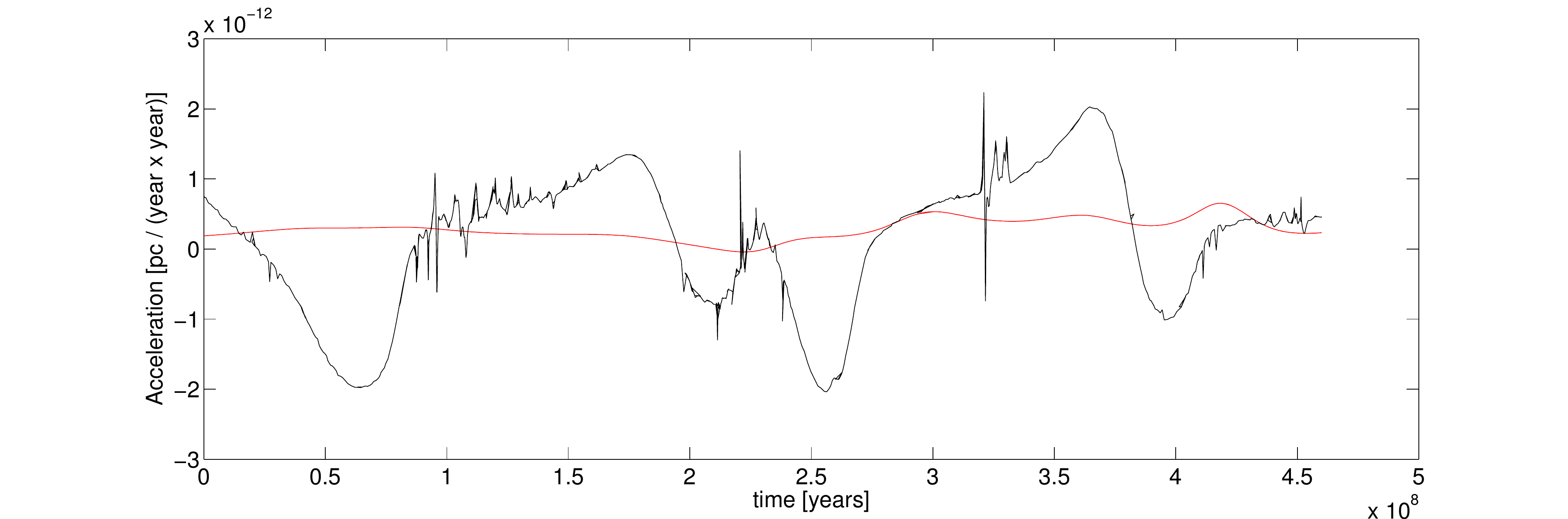}}
\caption{The varying radial forces per mass unit from spiral arms (black curve, low frequency variation), GMCs (black curve, overlaid high frequency variations) and
the Galactic bar (red/gray curve) as a function of time for 1/10 of the full simulation time range for a particular test particle. 
The mean distance of this particle from the 
Galactic centre is 6.8 kpc. The acceleration due to the overall-Galactic potential was about $2\cdot 10^{-11}$ pc\,year$^{-2}$. When calculating the 
comparatively little varying acceleration due to the Bar, the mean acceleration from a point mass of corresponding mass in the centre of the Galaxy has been subtracted.} 
\label{forces}
\end{figure}

We have generated a great number of sets of simulated orbits for test particles moving for 4.6 Gyr
with different parameters in the recipes for the various potentials and the number and distribution of GMCs.
Here, we concentrate on one homogenous set with varying the main contributors to the gravitational potential,
the overall Galaxy, the Spiral arms, the Galactic bar and the GMCs. 
Results of simulations with altogether
300,000 GMCs, are summarized in Table \ref{tab:model_results} and illustrated in  Fig. \ref{orbit1} and Fig. \ref{orbit7}. In the Figures, 
some typical orbits are illustrated in the $R-z$ plane, rotating
in the Galaxy model with the momentary angular speed so that the test particle stays in the plane. Fig. \ref{orbit7} shows an
orbit that ended at high latitudes. As is seen, the effects of the inhomogeneities in the potential, in particular those caused by
the GMCs, lead to considerable deviations from the standard  ``boxy-shaped" regular orbits shown in standard text books. 
The strengths and variations of the
different forces that attract the test particles radially in the Galactic plane are illustrated in Fig. \ref{forces}. 

In Table \ref{tab:model_results} resulting dispersions after 4.6 Gyr are given in
U, V and W velocities, i.e. in the $R,$ $\phi$ and $z$-directions, respectively, as measured in the Galactic plane at the distance of 8 kpc from the 
Galactic Centre. Here, all test particles around the Solar circle after 4.6 Gyr are included, not only the surviving cluster-representing particles. 
However, in order to get good enough statistics, we had to widen the ring to 
include all particles within a range of $7\,$kpc $<R<9\,$kpc, then compensating for the R-dependent differential rotation.
This, as well as the limited statistics, brought some errors into the dispersions. Based on a sequence of different simulations 
all including the effects of the Bar, giant molecular clouds and spiral arms, here denoted "BGS simulations", and 
realistic variation of the model parameters, we estimate that the relative errors may amount to $10\%$ in 
$\sigma_U$, $15\%$ in $\sigma_V$, while $\sigma_W$ is less affected. None of these errors affect our conclusions, but the values given
in Table \ref{tab:model_results} should not be overinterpreted.  In addition, local velocity dispersions are presented in the table for all the test particles,
thus mainly biassed towards the inner parts of the Galaxy as a result of the exponential density distribution of the disk. 
Also given in the table are measures of migration in the radial direction of the test particles, as well as the fraction $f_{400}$ of test particles that at 4.6 Gyr have $|z|> 400$ pc,
and the probability of survival of those when representing clusters, $S_{400}$, as following from Eq. (\ref{bcond}).

As is seen in Table 3 the 
inclusion of forces from the GMCs and spiral arms contribute essentially to increasing $\sigma_U$ in the inner Galaxy while spiral arms and the Bar
are less important for stars in the solar neighbourhood. For $\sigma_V$ spiral arms and, less so, GMCs 
contribute importantly and, as it seems, nonlinearly. For the velocity dispersion perpendicular to the Galactic plane, the dominating 
factor is the scattering by the GMCs. It should be noted that the heating effects of the Bar close to resonances may be of significance.

\begin{figure}[htbp]
\centering
\resizebox{\hsize}{!}
   {\includegraphics{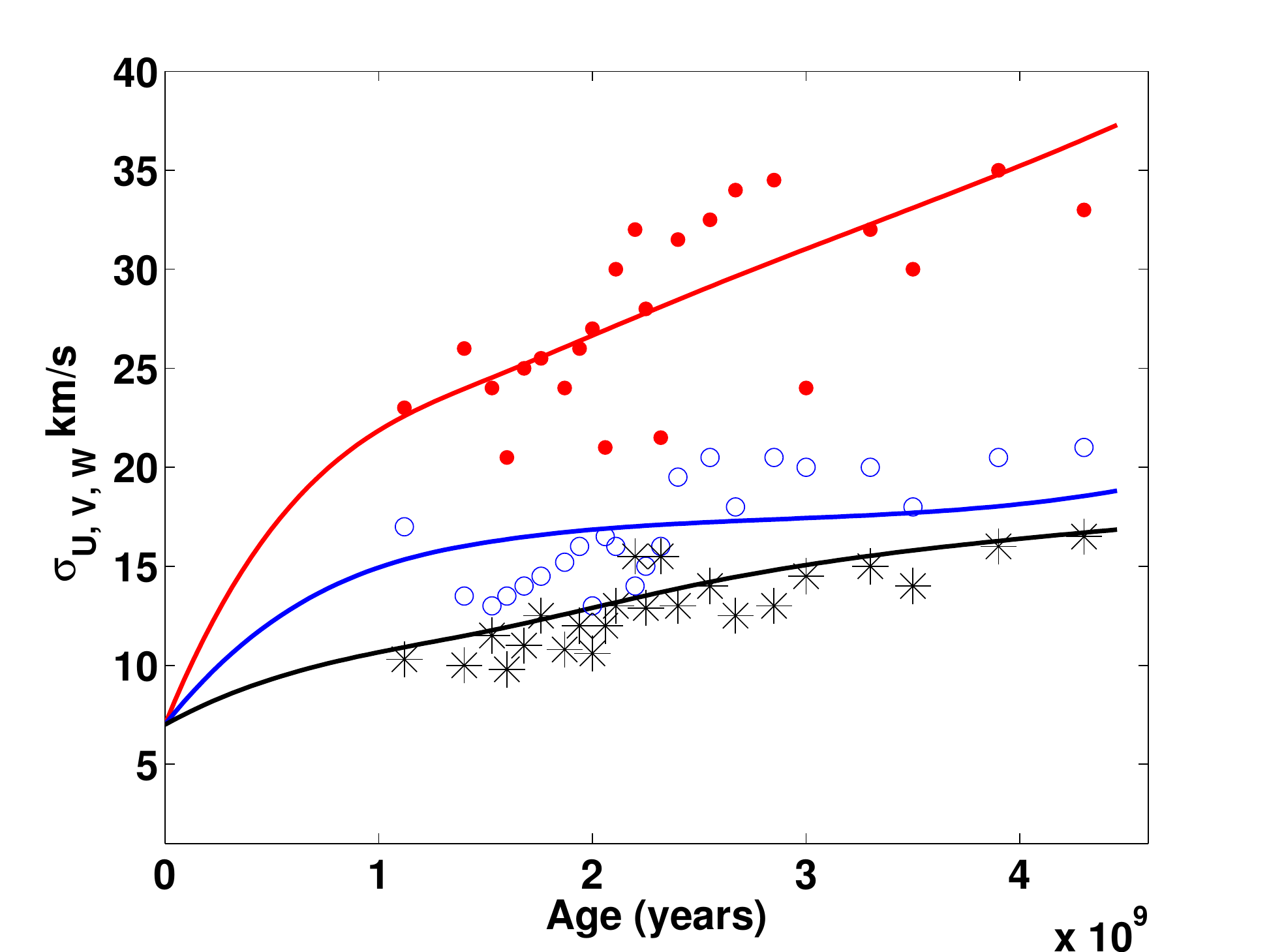}}
\caption{The dispersions $\sigma_{U8}$ (red lines), $\sigma_{V8}$ (blue lines) and $\sigma_{W8}$ (black lines) in U, V and W velocity components, respectively, 
calculated for stars around $R=8$ kpc in a BGS simulation (i.e. with Bar, GMCs and Spiral arms included), full lines, and when crossing the Galactic plane (for $\sigma_W$). Also plotted are the stellar 
observations of \citet{Holmberg09} as dots (U: full, V: open) and asterisks (W), with a colour coding corresponding to that used for the models.}
\label{dispersions}
\end{figure}

In Table 3, also the ratio $\sigma_U/\sigma_Z$ at the Solar circle is given. It is seen that the ratio stays safely below the critical value of 3.4 above which an
infinite slab will be subject to bending instabilities, see \citet{Sellwood96} and references therein.
In Fig. \ref{dispersions} we display the variation of the velocity dispersions as a function of time, compared with the observations of Solar-type stars of 
different ages in the Solar enviroment by \citet{Holmberg09}. As is seen, a rather good agreement with observations is found, in particular for $\sigma_U$ and $\sigma_W$. 
The calculated values for $\sigma_V$ are somewhat high for ages less than 2 Gyr. It is interesting to note that the calculated 
values of the dispersions, for models with N(GMC) increased from 300,000 to 460,000, increase by typically $10\%$ for the oldest stars and less than half of that
for the younger ones. A model with a higher value of N(GMC) would have led to a better agreement with the observed slope of the $\sigma_V$-age relation, although 
the absolute values of $\sigma_V$ would be too high. 

We see from Table 3 that while the test particles in the model galaxy as a mean have not moved substantially in the radial direction, the range of individual migration is
fairly extensive, with a dispersion {bf $\sigma_{\delta R}$} of about 1 kpc for models with GMCs and/or spiral arms. 
The stars of Solar age in the Solar neighbourhood, however, are predicted to have formed further in, at
a mean galactocentric distance $<R_0>$ typically 300 to 600 pc closer to the Galactic centre, essentially reflecting the asymmetric effects of scattering, due to the exponential 
star density in the Galactic disk. This scattering is provided by both the spiral arms and the GMC scattering. The differences
between the present Galactocentrick distance of the Sun and $<R_0>$ are somewhat greater than that 
obtained in simulations by \citet{Yasutomi91}, but smaller than that (of 2 kpc) suggested by \citet{Wielen96} for the Sun in an attempt to explain
its comparatively high metal content, as well as by \citet{Sellwood02} as a result of transient spiral arms, see also \citet{Bland-Hawthorn10} and \citet{Minchev12}. 

A glance at Table 3 might give the impression that the effects of stationary (though rotating) spiral arms on the velocity dispersions and on the migration of stars 
are relatively small in general. This impression is misleading, however:  the spiral arms contribute
substantially to the velocity scatter in U and V in the inner Galaxy, albeit still little to the scatter in W.  The differences between the spiral-arm effects at the
Solar circle and inside it may be ascribed to the general exponential profile of the Spiral arms and the fact that the co-rotation radius (at 9.16 kpc with the data adopted) 
is not so distant from the Solar circle. 

\begin{table}[h]
\caption{Results after 4.6 Gyr of model simulations for different Galactic models: BGS = Bar+GMCs+spiral arms; BG = Bar+GMCs, no spiral arms;  BS = Bar+spiral arms, no GMCs; $\Phi$ = Only overall Galactic potential.
Dispersions in $U$, $V$ and $W$ velocities for model stars around the Solar circle are given (indexed by ``8"), as well as the corresponding dispersions for all
the test particles (indexed by ``a"), means of the 
radial migration at the end of the simulation of all the stars from their starting points $<\delta R>$, the dispersion of these measures in the radial direction $\sigma_{\delta R}$, 
the mean original Galactocentric radius of all stars around the Solar Circle $<R_0>$ at 4.6 Gyr, the fraction of test particles ending with a distance greater than 400 pc
from the Galactic plane $f_{400}$, and the fraction $S_{400}$ of such particles (ending above 400 pc) that 
would survive were they clusters. The fractions given suffer from small number statistics.}
  \begin{tabular}[l]{lrrrr}
    \hline
Quantity & BGS & BS & BG & $\Phi$\\
$\sigma_{U8}$ [km/s] & 37.5 & 24.8 & 36.7 & 17.1\\
$\sigma_{V8}$ [km/s] & 18.9 & 14.0 & 16.1 & 7.7\\
$\sigma_{W8}$ [km/s] & 16.9 & 8.8 & 15.2 & 9.1  \\
$\sigma_{U8}/\sigma_{W8}$ & 2.21 & 2.82 & 2.41 & 1.88 \\
$\sigma_{Ua}$ [km/s] & 58.0 & 47.9 & 36.1 & 13.5 \\
$\sigma_{Va}$ [km/s] & 38.6 & 35.2 & 14.8 &  7.7\\
$\sigma_{Wa}$ [km/s] & 19.8 & 10.9 & 18.1 & 10.1\\
$<\delta R>$ [pc] & 74 & -256 & 195 & 85 \\
$\sigma_{\delta R} $ [pc] & 1161 & 1020 & 946 & 409  \\
$<R_0>$ [pc] & 7440 & 7700 & 7410  & 7830\\
$f_{400}$ [$\%$] & 1.8 & 0.2 & 1.8 & $< 0.2$  \\
$S_{400}$  & 2/6 & 1/1 & 1/9 & -\\

   \hline
  \end{tabular}
  \label{tab:model_results}
\end{table}

\begin{figure}[htbp]
\centering
\resizebox{\hsize}{!}{\includegraphics{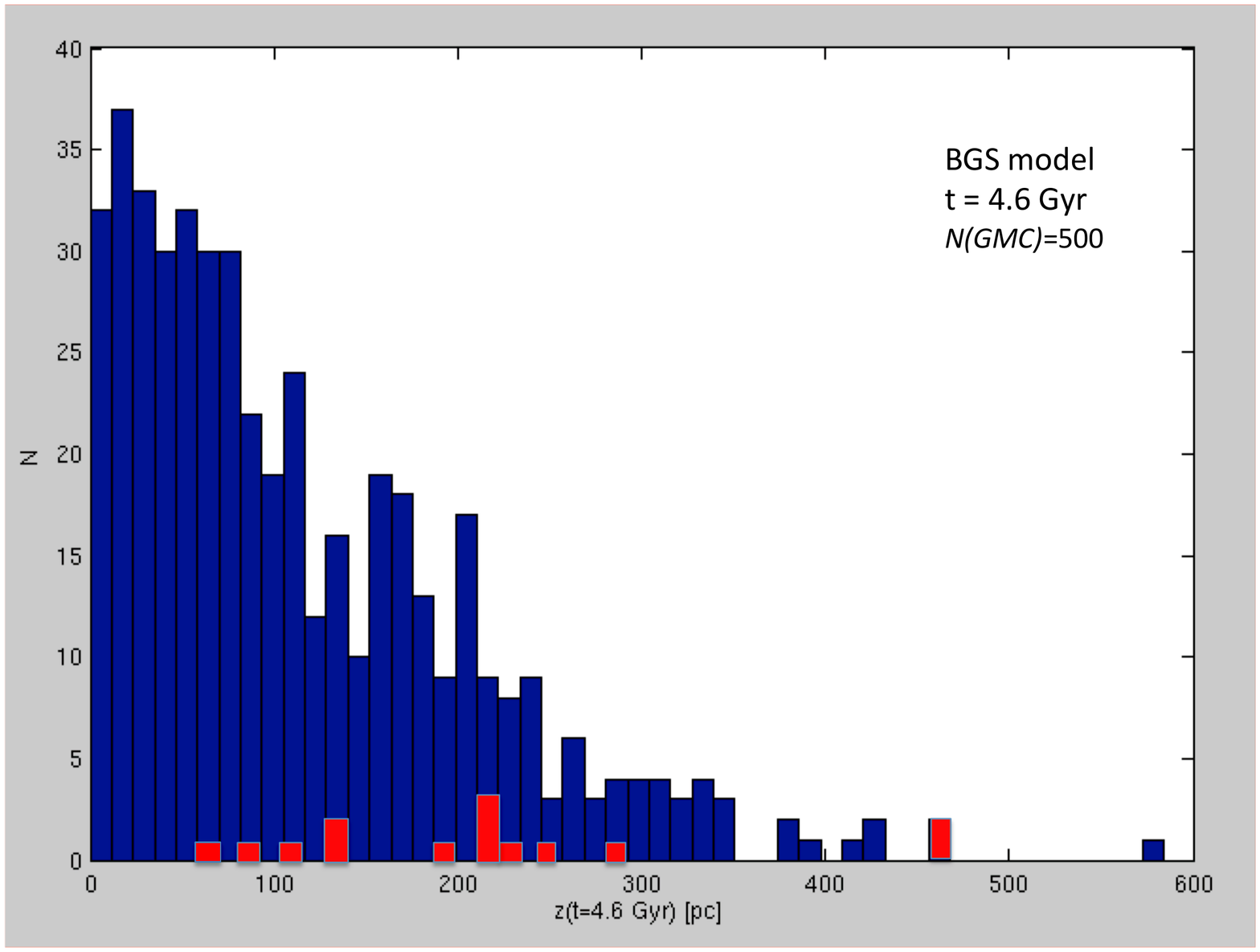}}
\caption{The $|z|$ distribution of test particles after 4.6 Gyr, in a BGS simulation with 300,000 GMCs. Marked in red are the model star clusters that would
survive for the full period, according to Eq. \ref{bcond}.}
\label{zdistr}
\end{figure}

In Fig.\ref{zdistr} the distribution of distances from the Galactic plane after 4.6 Gyr
for the test particles of the BGS simulation is displayed.
It is seen there and from Table 3 that there is a tail of high-latitude orbits resulting, but the fraction of orbits ending 
above 400 pc  is low, ranging in the interval $0.5-2\%$ for different runs, the exact value depending 
on the details of the details in the Galaxy model such as the number of GMCs and the $z$ dependence of the potential.

\section{The consequences of detailed GMC structures}

An importance circumstance to consider is to which extent the results of our simulations depend on the particular assumptions made concerning the structure of the GMCs.
The referee of the present paper made the interesting comment that the formation of the GMCs may occur at the expense of gas in their surroundings, such
that the scattering effects of the GMCs should be significantly reduced, as compared with our standard model where the GMCs are just added to the homogeneous disk
with the total disk mass (and thus gas density) correspondingly globally reduced. The GMCs may have formed out of the neighbouring gas such that a hollow region might result around it.
This hollow region, approximately a sphere, might produce a ``shielding" effect, such that the net gravitational force from the GMC felt by the test particle/cluster at a distance greater than the
radius of the hollow sphere (in practice
beyond 50 to 200 pc, depending on the cloud mass and the height above the Galactic plane), would then be unimportant. 

In testing these ``shielding" effects, we complemented the
standard GMCs in the BGS simulations by just adding an empty sphere, a ``void", centered around each of them, with a radius $R_v$ determined by the mass of the GMC and the mean density of the
surrounding disk. With such an approach one can easily prove that the ratio of the gravitational force from any individual GMC, relative to the standard case (when the GMC is just added
to the general gas background and the gas density is correspondingly reduced), $f_i$, is 
\begin{equation}
f_i = \left\{
\begin{array}{lc}
   \frac{\rho}{\rho_{red} } \alpha^{-2} \left\{1-\alpha^3 \right\}                               &  d < R_v\\
   0 & d \ge R_v.
\end{array}
\right .
\label{eqn:sphere_cond}
\end{equation}
Here, $d$ is the distance from the cloud centre, $\alpha=d/R_v$ and $\rho$ and $\rho_{red}$ denote the density in the gas disk in the Galaxy and the density reduced to consider the corresponding 
formation of GMCs, respectively, so that the total mass of the Galactic gas disk is the same in both cases. In this approximation, the disk is considered a homogenous medium (except for the GMCs), infinite
in all directions. In order to explore the effects of the finite extension of the disk gas in the $z$ direction, we approximate it with an infinitely thin disk with constant surface density, though with a 
hole surrounding the GMC with a radius again corresponding to the mass of the cloud, and find the
corresponding force fraction $f_{iz}$ to be supplementing Eq.(\ref{eqn:sphere_cond}) for the $z-$component of the force,
\begin{equation}
f_{iz} = 
\begin{array}{lc}
   \frac{\rho}{\rho_{red} } \alpha ( 1+\alpha^2) ^{-1/2}                          & d > R_v.\\ 
\end{array}                 
\label{eqn:disk_cond}
\end{equation}
While the "spherical shielding" of Eq. (\ref{eqn:sphere_cond}) should overestimate the effect of the shielding, the "cylindrical shielding" of Eq. (\ref{eqn:disk_cond}) (within the approximation of the disk as a relatively homogenous body with a clear
correlation between GMCs and hollow regions centered around them) should underestimate the effect. In Table \ref{tab:check_shielding} we present results of orbit simulations with the two different approaches, as compared with 
the standard BGS simulation. As is clearly seen, the effects of the spherical shielding on the fraction $f_{400}$ of stars reaching high elevations above the Galactic plane, as
well as on the velocity dispersions at the Solar Circle, are very considerable, while the consequences of ``cylindrical shieldning" according to Eq. (\ref{eqn:disk_cond}) are smaller. 

We performed another type of experiments in studying the effects of non-spherical structures of the GMCs: We divided each cloud into two spherical cores
and displaced them horisontally and symmetrically within the still spherical void, with the void centre at their common centre of mass. Both 
cores were modelled with Plummer spheres. We 
chose the distances between the cores to be 30 pc and 100 pc, respectively (``dumbbell30" and ``dumbbell100" in Table \ref{tab:check_shielding}). This type of configuration
was inspired by the fact that the observed GMCs often have an elongated and clumpy or even filamentary structure, see e.g. \citet{Blitz93}. We set the angular momentum
of each such pair of subclouds to zero, in view of the relatively low angular momenta found for GMCs by \citet{Imara11}. (An increase of their momenta to the specific 
momentum of their parent gas clouds due to differential galactic rotation was found to increase the scattering effects.) These systems act as quadrupoles outside the 
voids, with a gravitational potential decreasing with distance $d$ in proportion to $d^{-3}$. The resulting velocity scatters are shown in the Table. It is seen
that the scatters are smaller than for the ``displaced" or ``reservoir" cases, and for the standard ``naked" clouds (i.e. when the clouds are not enclosed in voids). (At the interpretation
of the results of these experiments one should note that all of the dumbbell bars were assumed to be initially directed towards the Galactic Centre. During the life
times of the GMCs, the angle between the bars and the Galactic radius increased gradually, in the direction opposite to the rotation of the Galaxy, to a maximum of about 70 degrees.) 

\begin{table}[tab:check_shielding]
\caption{The results of changing the representation of GMCs from
spherical clouds superposed onto an exponential gas disk model (the standard BGC case)
to models with clouds 
surrounded by voids with a subtracted mass from the gas disk equal to the mass
of the corresponding clouds. Data for different cases, with ``spherical shielding", ``cylindrical shielding",
clouds split into equal components with a mutual distance of 30 or 100 pc, respectively,
but with a centre of mass at the centre of the corresponding void (``dumbbell30" and ``dumbbell100"), 
cloud centres displaced relative to the correponding void (``cloud displacement"), and clouds built on the expence
of larger ``reservoir clouds" are presented.  
The velocity dispersions at the Solar circle in km/s and the fraction of stars 
with a distance of $>400$ pc from the Galactic plane at 4.6 Gyr are given. 
}
\centering
  \begin{tabular}[l]{lcccc}
    \hline
   Model   & $\sigma_{U8}$ & $\sigma_{V8} $ & $\sigma_{W8} $ & $ f_{400}\, [\%] $\\
   \hline
     Standard BGS &  37.5 & 18.9 & 16.9 & 1.8 \\
     ``spherical shielding" &  28.4 & 13.4 & 9.1 & $<$ 0.2\\
     ``cylindrical shielding"  & 33.1 & 14.8 & 12.9 & 0.5\\
     ``dumbbell30" & 30.0 & 15.6  &10.5 &$<0.2$ \\
     ``dumbbell100" &30.3 & 15.5 & 12.2 &$< 0.3$\\
     ``cloud displacement" & 40.8 & 15.7 & 18.4 & 2.2\\
     ``reservoir clouds" & 47.9 & 20.6 & 20.0 & 2.9\\
   \hline
  \end{tabular}
  \label{tab:check_shielding}
\end{table}

The assumption of a spherical void in the gas disk, centered around each individual GMC is, however, not very realistic. A first argument against that configuration
is that the density gradient in a stationary disk in itself implies an initial displacement of the GMC from the centre of the assumed spherical gas region from which it was formed, a
region which is here supposed to define the borders of the void, by characteristically 1 pc. (Here we neglect dynamical effects except for a gravitational contraction of the cloud.) 
Displacements of this order of magnitude should grow exponentially, due to differential
gravitational attraction from the disk, so that the GMC is moved parallelly to the disk plane a considerable distance 
towards the (assumed) spherical borders of the void within the lifetime of the GMC. 
In order to test the effects of
such displacements, we have carried out simulations with the clouds randomly shifted to distances $\Delta$ from the void centres with $0<\Delta<R_{v}$. 
The consequences of such ``cloud displacements" are seen in Table \ref{tab:check_shielding}. 
Basically, the scattering effects are due to the
fact that empty voids as such are effective scatterers, in the sense that the net local effects of the remaining gas in the disk are unbalanced if 
the gas in the void is removed. 
In general, the scattering effects of inhomogeneities in the disk, whether due to clumps of dense gas or corresponding holes in  the gas distrubution, 
may thus be rather similar. The effects of GMCs and local neigbouring or surrounding but off-centered voids may combine to significant scattering effects, producing wide scattering angles.
Outside the voids, these systems act as dipoles, with a potential proportional to "the dipole moment", mass$\times$separation between the centre of the void and the GMC, and the inverse
square of the distance from the system.

The degree to which local mass conservation prevails when GMCs form in the Galaxy certainly depends on the possible effects of non-circular and 
irregular mass motions in the Galactic field. We explored the effects of relaxing this condition locally by respresenting the Galactic gas disk between 4 kpc and 9 kpc 
from the Galactic centre as a whole by a number
of  288 overlapping flat spheroids, with half axes in the Galactic plane of 500 pc each, and half axes perpendicular to the Plane of 40 pc.  The density distribution
 within the spheroids was assumed to vary linearly in all directions from their centres, and the total mass was equalled to that of the total gas disk -- the corresponding 
potential was then subtracted from the general Galactic potential and replaced by the individual contributions from the spheroids. The spheroids were 
assumed to corotate in circular orbits around the Galactic centre with the local speed.
Each GMC was then located within at least one such spheroid at any given time, and the gas forming the GMC was subtracted from the total mass of the 
spheroid, keeping the shape of its linear mass distribution. Similarly, when the GMC was dissolved its mass was given back to the spheroid where the GMC was located. 
If it was located in two spheroids simultaneously, its gas contribution was shared proportionally between them.  The differential gas
velocities needed for a model of this type to be dynamically realistic must be on the order of 6 km/s which is acceptable in view of the observed velocity scatter in the gas (see Section 3.2 above). The results of this experiment are shown in Table \ref{tab:check_shielding} 
in the line denoted ``reservoir clouds". It is seen that this arrangement causes a great velocity scatter. 
Obviously, the additional inhomogeneities introduced by representing the gas disk in this way (which is in itself not grossly unrealistic, c.f. the observed 
maps in \citet{Nakanishi16}) adds to the scattering effects. 
 
Summarizing the results of the experiments with correlations between the GMCs and regions poor in gas (voids), supposed to represent the regions from which the
GMC gas originated, we find that a high geometric correlation is needed in order to severely diminish the velocity scatter and
spread perpendicularly to the Galactic plane from the results for the case with ``naked" clouds. However, it is questionable whether this type of correlation, 
between GMCs and gas-poor regions surrounding them is present at all in real galaxy disks. The distribution
of gas in galaxies has recently been explored in high spatial resolution, both observationally and theoretically. Thus, for $24-100 \, \mu$m dust emission from the LMC, \citet{Block10} found 
a structure with a power spectrum extending from $k^{-1}\sim 7.6 $ pc - $5000$ pc with essentially two power laws, $P \sim k^{\nu}$ with $\nu$ values of 2 and 3, the latter applying
for the interval $k^{-1}\sim7.6 $ pc - $200$ pc. This was well reproduced by hydrodynamic galaxy simulations by \citet{Bournaud10}. At a 50 pc resolution, the study by 
\citet{Druard14} of CO gas in M33 does not seem to show any correlation between dense regions and surrounding voids. Altogether, these results suggest
that high-density regions tend to be surrounded by regions with excess densities sooner than voids. We note that also studies of the distribution of young stars
in neigbouring galaxies suggest similar correlations of dense regions with moderately dense regions, sooner than with relatively empty regions (\citet{Gieles08}, \citet{Bastian09}, 
\citet{Bastian11}, \citet{Elmegreen14}).

We have studied the distribution of gas of our standard models, i.e with no voids and a smooth exponential gas disk, 
projected onto the model plane, and found a typical slope of the logarithmic power 
spectrum of about 2.2 in the wavelength range extending from
typical GMC radii to those of corresponding hypothetical hollow regions around them ($k^{-1}\sim 10 - 100$ pc). The corresponding slopes for models
{\it with} the hollow spheres implemented around them is about 2.0. Corresponding differences occur in the two-point correlation functions, which is
not unexpected, since they are directly related to the power spectra. From this, and the observed slope closer to 3 than 2,  
we conclude that reduction of scattering effects by GMCs by surrounding
voids is not expected to be important; more likely enhanced scattering effects may occur as a result of the piling up of gas around the GMCs. 

We have found that the fragmented structure of the clouds may significantly affect the gravitational scattering by the GMCs. Also, 
the time evolution of their structure is important. This evolution, influenced by the gravitational interaction of the cluster with the fragments, each of a mass perhaps comparable to that 
of a young massive cluster, is supposed to accelerate/decelerate the cluster but may also lead to scattering into increased angles. We have studied this possibility
by ``shooting" point-mass particles onto schematic models of GMCs, consisting of 4-10 diffuse blobs in mutual gravitational interaction, each with masses of
$10^5$ M$_{\odot}$ and with random velocities below the velocity of escape from the system. We then found the expected widening of the distribution of
scattering angles and a change of the particle velocity distribution, when compared with scattering by a homogenous cloud of similar total radius and mass. 
Although most particles are braked by the dynamical friction from the
cloud complex, as much as $10\%$ of the particles were typically found to increase their speeds at infinity from 10 km/s to 20 km/s or more. Also, the break-up of clusters by cloud
collisions may be severely dependent on the detailed cloud structure. More realistic studies of such effects, using 
detailed models of star-forming regions by Haugb{\oe}lle, Nordlund \& Padoan (cf. \citet{Padoan14}) are being pursued.

\section{Discussion and conclusions}
\subsection{The heating of the Galactic disk and the destruction of clusters}

Our simulations for test particles in a reasonably realistic Galactic potential suggest that at least for the last 5 Gyr, the observed 
heating of the thin Galactic stellar disk can be 
understood as the result of the effects of regular spiral arms, mainly accelerating the stars in the Galactic plane, and GMCs, 
linking the stellar orbits towards higher altitudes, a mechanism early on discussed  by \citet{Carlberg87}. In our simulations 
of the orbit evolution of star clusters, we have represented the clusters by simple
test particles, however modelling the break up of the clusters by tidal forces due to the 
Galactic mean potential as well the GMCs. The effects of the latter are quite important. 

The $z$ distribution of the test particles that would have survived as clusters according to the first and standard recipe
in the standard BGS simulation, is shown in red in Fig. \ref{zdistr}. We find there and from Table 3 that only a relatively small fraction $S_{400}$ of the M\,67-like clusters should
survive for 4.6 Gyr. We have studied the fraction of survivals as a function of the height $z_{end}$ above the Galactic plane in the end of the integration 
for a number of simulations and find the fraction to be typically a few percent for
$z_{end} < 100$ pc, rising to about 15\% at about 200 pc, 35-50\% at about 400 pc and next approaching even higher values.
It should be stressed that these estimates are based on small-number statistics and schematic; note for instance that clusters being close to the Plane but with high W velocities have a higher 
chance to survive individually.  For $N(GMC)$ increased to 460,000 the survival probability for the clusters that presently could be at heights above
$|z|=400$ pc is significantly reduced. This reduction is, however, roughly compensated for by a higher fraction of
particles scattered to those heights, so that, in the end, a percentage $f_{400} \times S_{400} \sim 0.5\%$ of all massive open clusters formed are predicted to end with $|z|>400$ pc, rather independently
of $N(GMC)$ for the range $300,000<N(GMC)<460,000$. Within the basic assumptions of the present Galaxy BGS model we find that the series of simulations suggest that this percentage
has statistical standard errors of $\pm 0.1\%$. As can be seen from the last column in Table 4, the systematic errors due to our limited understanding of the internal GMC structure
could be considerably greater.

\begin{figure}
\resizebox{\hsize}{!} {\includegraphics{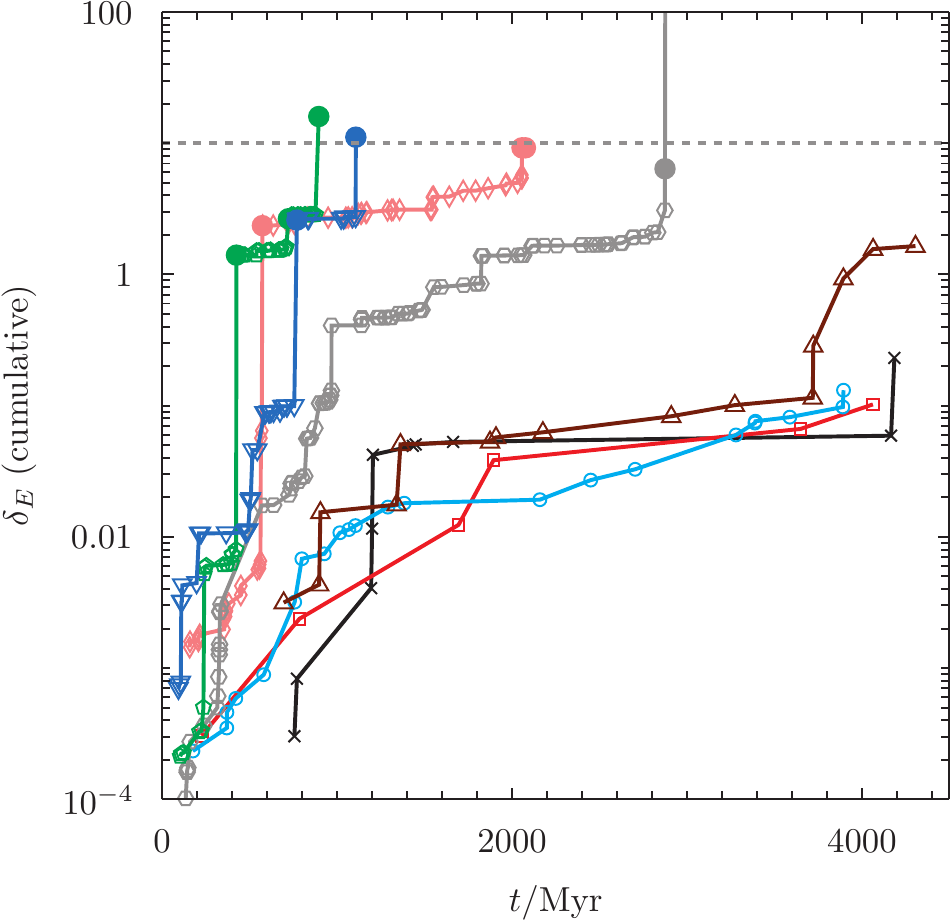}}
\caption{The history of the energy absorbed by clusters in our synthetic cluster
encounter simulations, $|\delta_E|$, as a function of time $t$.  The clusters
shown are the eight in total whose ultimate heights above the Plane
$|z_{end}|>350\,{\rm pc}$.  Each line shows a separate cluster, marked with a
different point type and colour.  Each point is an encounter between a cluster
and a GMC.  Large, filled, round points are those where $|\delta_E|>1$ for a
single encounter.  The dashed grey line marks a cumulative  $|\delta_E|$ value
of ten, above which clusters are not found to survive.}
\label{fig:highzdE}
\end{figure}

\begin{figure}
\resizebox{\hsize}{!}{\includegraphics{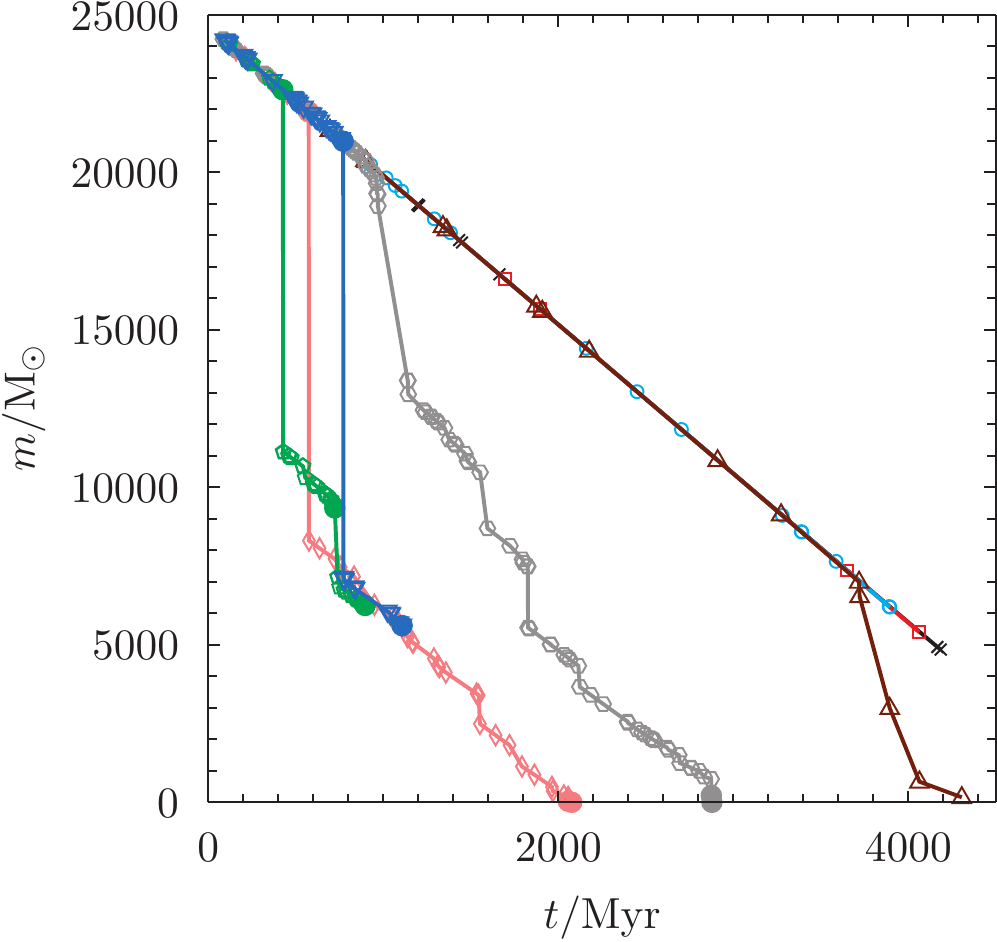}}
\caption{The history of the stellar mass $m$ in clusters in our synthetic cluster
encounter simulations as a function of time $t$.  The clusters shown are the
eight in total whose ultimate heights above the Plane $|z_{end}|>350\,{\rm pc}$.  Each
line shows a separate cluster, marked with a different point type and colour.
Each point is an encounter between a cluster and a GMC.  Colours and point types
are the same as for the clusters in Fig.~\ref{fig:highzdE}.}
\label{fig:highzm}
\end{figure}

We have checked our rough standard procedure, according to Eq. (\ref{bcond}), by applying the more detailed 
treament of Sec. 3.2. The survival of clusters at high latitudes turns out to be relatively independent of
whether we use the criterion of Eq.~(\ref{bcond}) or the analysis based on $N$-body
simulations described in Section 5.2.  Fig.~\ref{fig:highzdE} shows the
internal energy absorbed by all eight particles that are scattered up to
$|z|>350\,{\rm pc}$ at $4.6\,{\rm Gyr}$ in a model simulation 1000 test particles.  
In this particular calculation the clusters were all considered to have an
initial half-mass radius of $1\,{\rm pc}$.  Four of these clusters undergo
strong scatterings, which lead to vertical lines in the plot.  This causes them
to become rapidly disrupted thereafter, as can be seen by comparing the rate of
loss of mass from the same clusters in Fig.~\ref{fig:highzm}.  Four of the
clusters do not suffer strong scatterings and hence survive until late times.
One is destroyed by a series of moderate-intensity encounters, but the other
three survive to an age of $4.6\,{\rm Gyr}$.  Similar results are obtained for
larger initial half-mass radii.

\begin{figure}
\resizebox{\hsize}{!}{\includegraphics{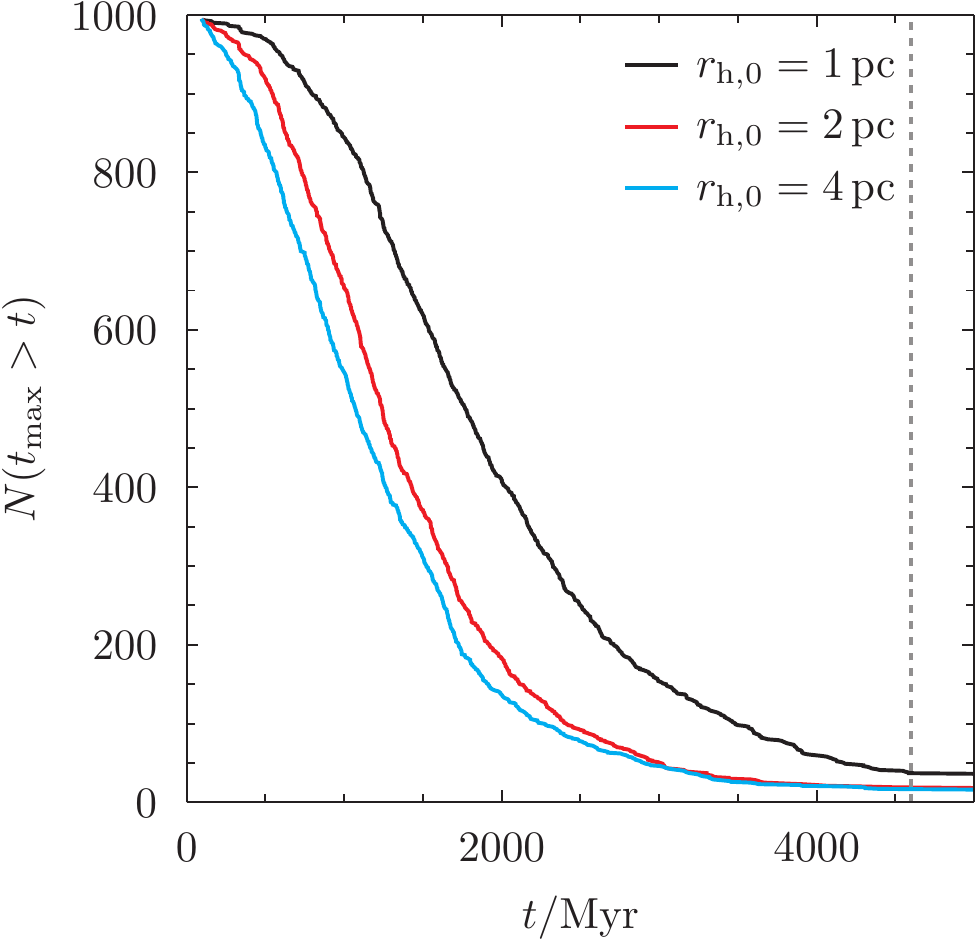}}
\caption{The number of clusters $N$ surviving for at least time $t$ in
a BGS simulation.  The black, red and blue lines show initial
cluster half-mass radii of $1$, $2$ and $4\,{\rm pc}$.  The grey dashed line
marks $4.6\,{\rm Gyr}$.}
\label{fig:cumulativeLifetimes}
\end{figure}

The distribution of lifetimes in one of our BGS simulations with 1000 test particles/clusters, following
the synthetic ($N$-body) cluster destruction model, are shown in
Fig.~\ref{fig:cumulativeLifetimes}.  The median lifetime decreases with
increasing initial half-mass radius, as the clusters are larger,
more heated by encounters with GMCs.  The median life time for 
$r_{h,0}=1$ pc clusters is almost 2 Gyr which departs from the estimate given by 
\citet{Wielen71} that only 2\% of the open clusters survive for more than 1 Gyr. The dominant
reason for this is that our model clusters are at least one order of magnitude more 
massive than typical Galactic open clusters. Only a small number of clusters, however,
survive until $4.6\,{\rm Gyr}$ (marked with a dashed line on the plot). 
For an initial half-mass radius of $1\,{\rm pc}$ the number of surviving clusters is close to
that found using Eq.~(\ref{bcond}).  In fact, all 36 clusters
in the simulation that survived in the approximate treatment also survive in the $N$-body approach; 
the latter approach just adds one more cluster to this category. This agreement may seem
astonishing, since the simulation was carried out with the standard choice of 
a cluster radius of 3 pc in Eq.~(\ref{eq379}) while the initial $r_h$ value in the
$N$-body simulations is 1 pc. However, $r_h$, representing the median radius, is supposed to 
be smaller than the radius used in the derivation of Eq.~(\ref{bcond}) and moreover, as is illustrated in Fig.\ref{fig:fitRh},  
$r_h$ varies in the $N$-body simulations and is at all times greater than its initial value. 
The principal difference between the two
formalisms is that the $N$-body treatment takes into account the variation of
half-mass radius over the lifetime of the cluster.

There are shortcomings of the
Rutherford (impulse) scattering approximation in treatments of the GMC-cluster interaction when a Galactic potential.
Yet, this approximation is shown to give  a good estimate of the critical impact parameter for the destruction of 
clusters, as we have demonstrated by $N$-body simulations of the cluster destruction process.

\subsection{The origin of high-altitude old open clusters}

We shall now address the question whether it is reasonable to assume that the massive high-altitude old open clusters like
M\,67 and NGC\,188 could have originated in the Galactic plane and subsequently been brought to their high altitudes by scattering by GMCs,
as well as interaction with spiral arms and the Galactic bar. 

With the fraction for these clusters to be brought up to at least 400 pc, $f_{400}$, of about 1.8$\%$ as a typical value from our standard BGS simulation, 
and a probability of survival for such clusters against tidal break-up, including GMC collisions, of $S_{400}$ of typically 0.3, 
we find that the theoretical joint probability $f_{400}\times S_{400}$ for massive open clusters formed in the plane of the
Galactic disk to be located after 4.5 Gyr at $|z|>400 $ pc, is about 0.5\%. This is in fair agreement with the expected fraction of all formed
massive Galactic clusters to be found at such high levels, as estimated to {\bf F}$_{\rm obs}=0.2-0.5\%$ in Sec. 2.
It should be noted here that the estimates in Sec. 2 were made after the model calculations of Sec. 5 had been completed; thus the empirical result
did not at all affect the setup and constraints of the simulations.
We have thus found consistency with the hypothesis that M\,67 has its origin close to the Galactic plane in a normal
disk orbit. As regards the significantly older NGC\,188 its existence at an even higher altitude 
also seems consistent with the disk-origin hypothesis, and this may also be the case for the younger clusters NGC\,2420 and NGC\,7142, as well.

\begin{figure}[htbp]
\centering
\resizebox{\hsize}{!}
   {\includegraphics{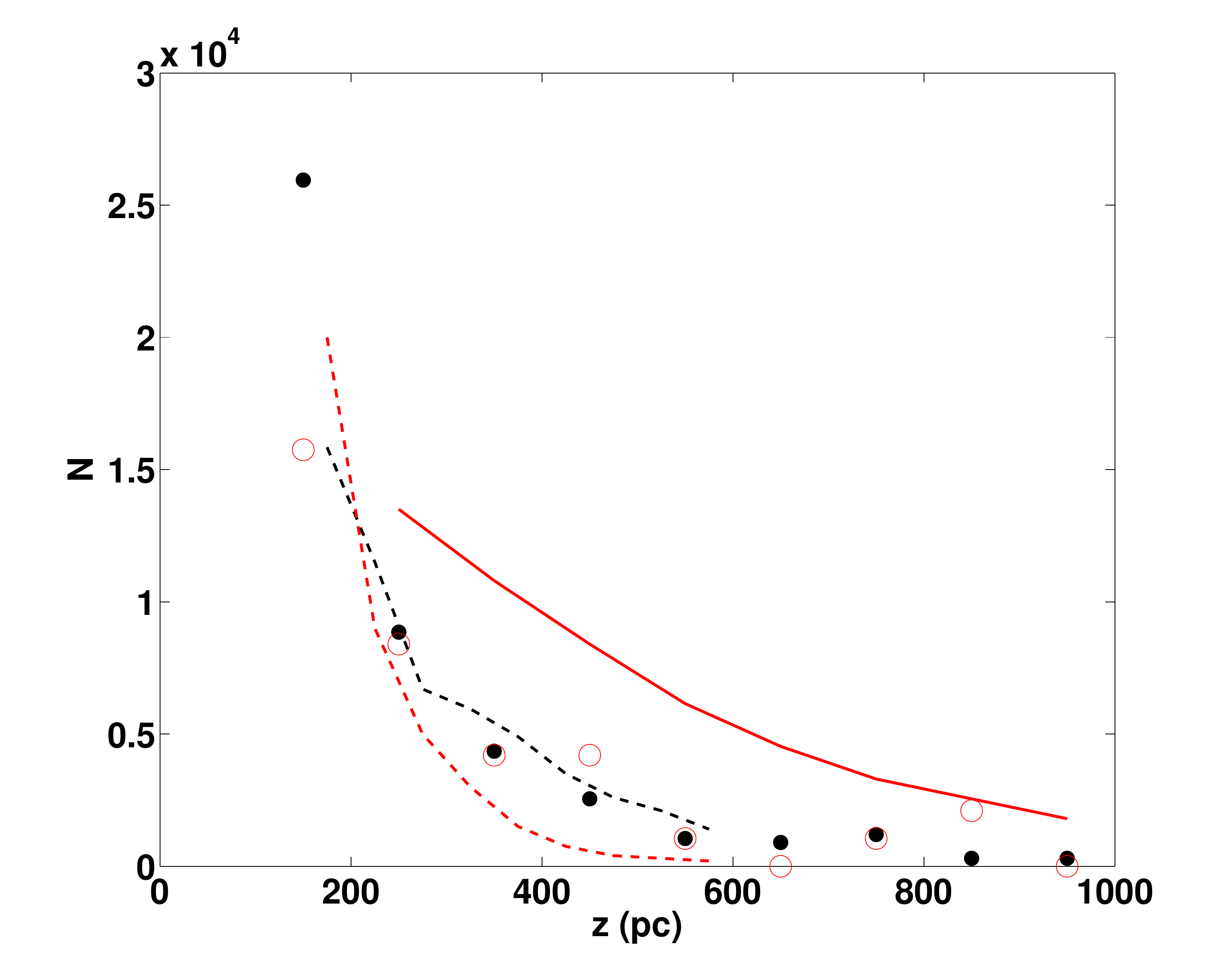}}
\caption{The predicted z-distribution of massive open clusters after 4.6 Gyr, as based on a BGS run with 1000 test particles subject to 
destruction due to GMC collisions (black dashed curve) and the corresponding distribution of stars (red dashed curve). The observed distribution from the WEBDA data base
(\citet{WEBDA}) is marked with 
black dots., and the distribution of old open clusters from \citet{Janes94} within a Galactic cylinder with radius 3 kpc is marked with red circles. The bump around
$|z|\sim 850$ pc in the latter data corresponds to altogether 2 individual clusters.
The full red/gray curve denotes the observed stellar distribution, according to \citet{Yanny13}. Note that the scale of 
the N axis is ambigous, a relevant comparison is the ratio of the number of clusters around, say, $z=200 $, as compared with that at
500 pc.}
\label{clustz}
\end{figure}

Some verification of the present scenario for the formation of clusters, presently at relatively high altitudes, can also be found by
comparing the distribution of the calculated heights z at 4.6 Gyr in Fig \ref{zdistr}, corrected with the probability of survival $S$. We find a relatively good agreement with the observed cluster distribution from the WEBDA data base as 
is seen in Fig. \ref{clustz}, as well as with the distribution of the open clusters of ages similar to that of the Hyades or older from the work by 
\citet{Janes94}, provided that we restrict the selection from their list of 72 clusters to those 42 that are within a Galactic
cylinder with the Sun on its axis and a radius of 3 kpc. This restriction was applied in order to avoid the strong bias in their sample
favouring high-altitude clusters at great distances, see the discussion in Sec. 2.  The limited statistics makes it unclear whether the bumps in the Janes \& Phelps distribution, when compared with our simulations, indicate that some extra heating phenomenon
or even ``unusual" formation mechanisms, like the ones discussed in Appendix A.2, are needed
to explain any of these high-altitude clusters. This comparison should be extended 
to a sample of cluster models with a distribution in initial cluster mass and age. 

In Fig. \ref{clustz} we have also plotted the observed distribution of stars, as well as the calculated z (again after 4.6 Gyr)
for our test particles, uncorrected with the $S$ factor, which should then represent the distribution of stars of that age. It is seen that the observed
stellar distribution is much broader -- most probably due to the fact that many stars at high latitudes are significantly older than 4.6 Gyr, and this population also includes 
thick-disk stars. 

Further simulation along the present lines, extending through the full life-time of the Galaxy, and more comparisons with more complete surveys of clusters 
of different ages, metallicities and at different locations, and not the least with coming Gaia data for both clusters and stars, would be valuable. Also, the dependence of
the present results on the structure of the GMCs and their surroundings stresses the need for improved simulations with more detailed physical
models of the insterstellar gas. 

\subsection{Did the Sun originate in M\,67?}

The possibility that the Sun was once formed in M\,67 was discussed by \citet{Onehag11} and \citet{Onehag14}, on the basis of their result that stars in M\,67 have an abundance pattern more similar to that of the Sun than most Solar twin candidates explored by 
\citet{Melendez09} and \citet{Ramirez09}. If the early M\,67 really would have been the original environment of the Sun, the arguments by \citet{Pichardo12} suggest that 
the Sun should have been formed in the cluster when it still had an orbit much closer to the Galactic plane
 and diffused out of the cluster. In some latter scattering process against one or several massive objects the cluster was then accelerated into its present high altitude orbit.
In this paper we have demonstrated that this scenario is possible, in the respect that the cluster may well have formed in the
plane and later been scattered up to its present locus by collisions with one or several Giant Molecular Clouds, perhaps pre-accelerated 
in the plane by spiral arms. 

There are, however, severe objections against this hypothesis. 
The first objection is based on the fact that the Sun, and the M\,67 stars explored by \citet{Onehag14}, are not unique -- several
carefully studied examples exist (see e.g. \citet{Melendez12}, \citet{Melendez14a}) which have remarkably Solar-like abundance profiles. \citet{Melendez09} note that a minor fraction (about 10-20\%) of their Solar twins from the Solar neigbourhood
resemble the Sun more than the majority of all twins (see also \citet{Nissen15}). It thus seems possible that the number of stars with the Solar abundance characteristics in the Galactic field is too great for all of them to come from one cluster. The detailed studies of the binary stars 16 Cyg (\citet{Tucci14}) and XO-2 (\citet{Ramirez15}) should also be noted in this connection.
In these stars, the components with discovered planets show a more solar-like abundance profile than their companions. The hypothesis brought forward by \citet{Melendez09} that the abundance pecularities of the Sun relate to its role as a planetary-system host, and not its more or less dense formation cluster, is supported by those findings.

The second objection arises from the expectation that the Sun left M\,67 fairly early in the evolution 
of the Solar System. There in one extra reason for such an early escape to be most probable, 
apart from the fact that it would give more time for the 
dynamical processes that we have discussed to bring M\,67 into its current, highly inclined orbit.
This reason relates to the existence of the Oort Cloud. In case the Sun was born in 
an M\,67-like cluster, it has been shown by Nordlander et al. (2016, in prep.) that any primordial 
Oort Cloud (i.e., one that was formed in connection with the scattering of planetesimals during 
the formation of the giant planets) would become disrupted in the course of some hundred Myr 
due to encounters with other cluster stars. This speaks in favour of a delayed Oort Cloud formation in connection with the Late Heavy 
Bombardment (LHB, \citet{Brasser13}). Ongoing work on the same mechanism, when 
placed in the framework of M\,67 or a similar dense solar birth environment (Rickman et al 2016, in prep.), indicates a very efficient comet 
emplacement into the Cloud; however, any prolonged dwelling of the Sun within the cluster 
afterwards would cause a threat to the survival of the Cloud.
Of course, if the Sun left M\,67 before the LHB, the picture of Oort Cloud formation by \citet{Brasser13} is also fairly attractive. 

Anyhow, an early exit implies that the Sun and its birth cluster have revolved many times in their Galactic Orbits with 
somewhat different speeds. This raises our third objection against the hypothesis of an M\,67 origin of the Sun: 
It would then be as probable that the Sun originated from any other of the M\,67-like clusters that were suggested above to be 
around, and even much more probable that it would come from any of the other similar but more numerous clusters closer to the Galactic disk, a cluster which is since long dissolved. 
The only reason for believing that the Sun
really has this specific origin would then be the similarity in the detailed chemical composition. It is important to study the chemical composition of other similar clusters to see if the
specific solar composition profile is
a characteristic signature of dense star-formation regions in general.

\begin{acknowledgement}
Johannes Andersen, Martin Asplund, Axel Brandenburg, Bruce Elmegreen, Gerry Gilmore, Ulrike Heiter, Oleg Kochukhov, Andreas Korn, Henny Lamers, Hal Levinson, Per Olof Lindblad, Jorges Mel\'{e}ndez, Poul Erik Nissen, Thomas Nordlander, \AA ke Nordlund, Hugo Pfister and Erik Zackrisson are thanked for valuable discussions and suggstions. The anonymous referee is thanked for several interesting points and valuable suggestions.
Nordita is thanked by BG for hospitity. This work was supported by the Swedish Research Council, grant numbers
  2008-4089, 2011-3991, 2012-2254, and 2012-5807.  
The simulations were performed on resources provided by the Swedish
National Infrastructure for Computing (SNIC) at UPPMAX and LUNARC, using hardware
funded in part by grants from the Royal Fysiographic Society of Lund.
HR was supported by Grants 74/10:3 of the Swedish National Space Board and 2011/01/B/ST9/05442 of the Polish National Science Center

\end{acknowledgement}

\begin{appendix}

\section{The possible formation of a M\,67-like cluster at high latitudes}

 In this Appendix
we shall explore whether the present location of the metal-rich high-altitude clusters directly discloses their mode of formation, and not a secular evolution of their orbits 
in the Galactic disk - this latter aspect is discussed in Sections 3-5.  
First we investigate the possibility that such a cluster formed due to relatively
normal star formation in an interstellar cloud belonging to the Galactic disk but in orbit with a considerable inclination to the Galactic plane, and next proceed to more ``unusual" formation scenarios. 

\subsection{Formation from gas in high-inclination orbits}
We can explore the possibility of star formation in high-inclination orbits by counting the number of known young stars at high altitudes, if we make the assumption of a not drastically
varying star-formation rate with time during the last 4-5 Gyr. \citet{Knude97}, based on the  survey by \citet{Oja92} of stars within 20 degrees from the Northern Galactic pole and complete for V$<11.5$ mag., explored the number density of stars of spectral type A2-A7 as a function of distance from the Galactic plane. From that we find a space density at $z=450$ pc of $1.4 \cdot 10^{-6}$ pc$^{-3}$ for the A-type stars with ages $\leq 0.75$ Gyr, while for the group of A-type stars with ages in the range 0.75 Gyr to 1.7 Gyr the corresponding density is $2.3 \cdot 10^{-6}$ pc$^{-3}$. Converting the spectral-type range to the mass interval $1.9 - 2.2$ M$_\odot$, following \citet{Torres10}, one finds from the IMF of \citet{Kroupa02} that this should correspond to a density of $2.6\cdot 10^{-5}$ stars pc$^{-3}$ earlier than spectral type M
or $19\cdot 10^{-5}$ pc$^{-3}$, if also M dwarfs down to a mass of 0.08 M$_{\odot}$ are included for the youngest age group. For stars in the older age range the corresponding number is {\bf $4.4\cdot 10^{-5}$} stars pc$^{-3}$ 
or $31\cdot 10^{-5}$ pc$^{-3}$, respectively. 
(It should be noted that we have here assumed all the stars of \citet{Knude97} to be main-sequence stars, i.e. we have neglected the contribution to his number-density estimates of blue horizontal-branch stars and other post-main-sequence stars, and of subdwarfs). Now, again guided by the estimates of the history of the star formation rate in the disk of \citet{Just11}, we assume a two times larger star formation rate 4.5 Gyr ago than presently at a height of 450 pc. We then find a maximum star formation rate of $\leq 6 \cdot 10^{-4}$ pc$^{-3}$ Gyr$^{-1}$. Thus, in order to provide the roughly 20,000 stars of a cluster like the initial M\,67 cluster for every Gyr, one would then need to take all stars within a box of 0.5 kpc $\times$ 0.5 kpc $\times$ 100 pc that could possibly have been formed in an M\,67-like orbit
for 1 Gyr into consideration. Indeed, if our cluster is part of a mass distribution of clusters, so that, as discussed above, the total contribution of this (now dissolved) cluster distribution would amount to 500,000 stars, this population, now being represented by our single cluster, would then correspond to the total star production of a box on the order of 2.5 kpc $\times$ 2.5 kpc $\times$ 100 pc. This demonstrates that the assumption that the M\,67-like clusters are just representative of the star formation in original high-inclination orbits, or even on high latitudes initially, seems not realistic if this star formation occurred at a rate proportional to the star formation rate in the Galactic disk today. In practice, a strong peak in star formation rate at high altitudes would have to be invoked, corresponding to an increase in this rate by about one order of magnitude. Thus, if the high-altitude clusters are just the result of normal star formation in the Galactic disk and not of such a considerable peak in the formation rate some Gyr ago, they should, as the other Solar-type stars at high altitudes, have experienced a significant orbit evolution since the formation time, bringing them to greater heights above the Galactic plane, i.e. by disk heating. 

Similar results are obtained if one considers the B-type stars at high altitudes. In his survey of stars within 20 degrees of the Northern Galactic pole and complete for V$<11.5$ mag.,  \citet{Oja92} found 23 stars in the spectral-type interval B0-B8. Of these, most lie beyond 500 pc. From these we estimate, in a way similar to that for the A-type stars above, adopting an interstellar extinction $A_{\rm B}$ of 0.077 mag., following \citet{Knude96}, a corresponding total star formation rate of $\leq 10^{-4}$ pc$^{-3}$ Gyr$^{-1}$.  Similar results were also obtained from the
magnitude-limited spectroscopic survey of early-type hydrogen rich stars at high Galactic latitudes of \citet{Saffer97}. 
 
We also note that \citet{Allen04} have studied the samples of O and B stars found at high Galactic latitudes by \citet{Conlon92} and others, and argue, on the basis of orbit calculations, that almost 90\% of them may be run-away stars from the Galactic plane (see also \citet{Tobin91}). This would then even more stress the need to invoke either a very high peak of star formation leading to stars at high Galactic latitudes
about 4 Gyr ago, or alternatively to argue that the orbits of the high-altitude metal-rich clusters have indeed been considerably affected by disk heating.

These arguments show that if the high-altitude clusters were formed in orbits close to their present ones, they were the result of an unsusual event in the star-formation history of the Galaxy, at least if the present state of star-formation is assumed to be rather normal. 

\subsection{``Unusual" formation scenarios}
Several ``unusual" formation scenarios have been suggested for may be argued for the metal-rich high-altitude
clusters; see the summary
by \citet{VandePutte10}. Several of these may, however,  be excluded in the case of the clusters in Table 1, primarily on the basis of their comparatively high metallicity. The following possibilities remain and should
be further explored:

(1) \citet{Martos99} modelled the gas response to the spiral arm density wave in a thick magnetized Galactic disk and found that the spiral shock could bring gas from the inter-arm region to high altitudes, followed by star formation, albeit with a low efficiency. 

(2) \citet{Friel95} suggested that impact of high velocity clouds on to the Galactic disk could lead to star formation in the Disk gas, a mechanism studied by \citet{Comeron92}, who related an oblique collision of this kind to the formation of Gould's belt. The cluster formed in the process could retain some kinetic memory of the event,  leading to a high z and possibly eccentric Galactic orbits. This interpretation of Gould's belt is, however, not generally accepted (for other alternatives, see, e.g., \citet{Lindblad20}). It is not clear whether such high metallicities as that of M\,67 could result in this case; \citet{Wakker99} found that High Velocity Clouds have low metallicities of about 0.1 Solar. 

(3) Globular clusters may impact the disk and cause disk-gas compression, either due to gravitational focussing (initially discussed by \citet{Wallin96}, who concluded that such passages, happening about once every $10^6$ years, may lead to a minor part of the star formation in the Galaxy) or shock wave formation, \citet{Levy00}. \citet{VandePutte09} have tried to trace such effects caused by recent passages of globular clusters through the Galactic disk, with no very significant results. This would presumably produce clusters with metallicities representative of the disk, and perhaps high-inclination Galactic orbits with considerable eccentricities. 
\end{appendix}

\begin{appendix}
\section{The numerical representation of the Galactic force field}
The general Galactic force field has been pre-calculated from mass points in a spatial rectangular grid, representing the mass distribution adopted from Model 1 of Binney (2012).
The general mesh size was 
50 pc, though with a decrease of the size in z (perpendicularly to the Galactic plane) to 10 pc for $|z| < 120$ pc and increase to 200 pc for $|z] > 2000$ pc. 
The forces in radial ($R$) direction, parallel to the plane, and in $z$ were calculated along lines at equal distances between the mass points in the mesh in order to avoid spurious variation as a result
of a force point close to a point in the mass grid.  The force was tabulated for every 200 pc in $R$ and every 50 pc in $z$, including one line
in the Galactic plane. In these force tables, we interpolated using cublic splines. The resulting forces were carefully analysed as a function 
of distance and height above the plane, to avoid wiggles. We also tried, as an alternative, representation of the force components by polynomina 
in $R$ and $z$ (up to fourth degree), derived from the tables of 
Allen $\&$ Markos (1987), and found a quite satsifactory agreement with the results from the table representation. In particular, there were no tendencies for the resulting scatters in U, V and W
to depart significantly from those obtained by interpolation in the force tables. We conclude that no additional velocity scatter was introduced as
a result of the interpolations.  

As a standard, the forces from the different
spheroids used in representing the spiral arms and the central Galactic bar 
were pre-calculated as a function of $r$ and angle $\beta$ between the symmetry axis and direction to the point in question,
and scaled with the squared distance from the centre and the cosine/sine of the angle, and next pretabulated as a function of distance and angle, for cubic spline interpolation
in the orbit calculations. In the pre-calculation, a cubic grid representation of the mass distribution was used, with a mesh size of 2.5 pc. The mass at each mass point was
then chosen to satisfy the density variation adopted for the spheroid. As an alternative, a Monte-Carlo approch was tried with $N$ homogeneous spheres filling the spheroid, each 
sphere with a typical radius of $R_s N^{-1/3}$, where $R_s$ is the  maximum size of the spheroid and the mass of each sphere was again chosen to satisfy the density
variation adopted in the spheroid. Here, $N$ was typically chosen to $10^8$. In either case, the tables generated 
in $r$ and $\beta$ were typically $200\times 50$ in size. In addition to the comparsion between the grid representation and the Monte-Carlo approach which showed 
satisfactory agreement, on the suggestion of the referee the method was checked in several ways. The appropriateness of the 
Monte-Carlo-representation was checked by some test cases relative to the formulae by Schmidt (1956), with high-order gaussian quadatures to evaluate the integrals. An excellent 
agreement was then found. In an attempt to 
check the spline interpolation, and in particular the risks that wiggles in the interpolation could contribute to the scatter in resulting velocities, we approximated the tables from the 
pre-calculations by polynomina in $\beta$ and $r$ or $1/r$, for points inside and outside the spheroids, respectively. These polynomia were scrutinized to avoid wiggles.
For points close to the spheroid surface, terms in both $r$ and $1/r$ were used, and special care was excercised in tying the approximations continuously together.  
Typically, $5\times5$ terms were necessary to approximate the tables with an accuracy on the order of $2\%$; an exception, however, was the force from the ellipsoids representing the 
spiral arms where both force approximations with both $r$ and $1/r$ terms were needed and the forces of the pre-tabulated tables were represented with maximum errors of 8\%. 
When alternatively applying these polynomina instead of the standard interpolation in the tables, the resulting differences in calculated velocity scatters for the test particles were found to 
be less than 2 km/s. Yet, the errors in calculating the forces from the spiral-arm spheroids seemed worrying, since the test particles/clusters in our simulations interact closely
with them, and even may pass through them, with an abrupt change in the force derivates when the test particles enter from an assymptotic $1/d^2$ to a $d$ dependence, $d$ here representing
the distance from the centre of the spheroid. 

In order to test this further we chose to apply our pre-calculation and interpolation techniques to {\it spherical} clouds, for which, alternatively, the
forces could be calculated analytically, in our Galaxy model.
We performed these experiments with only 200 test particles and a reduced number of model GMC:s ($N(GMC)=10,000$)
in order to save computer time; the
reduction of $N(GMC)$ is expected to exaggerate the effects of the numerical errors in the handling of the spiral spheroid forces since the test particles are not scattered so efficiently to
high latitudes and thus spend more time close the Galactic plane, interacting with the spiral arms. The results are displayed in Table \ref{tab:check_spheroid}. 
When replacing the spheroids in the spiral-arm representation with spheres with corresponding masses (again with a linear variation with
radius of the linear density and with radii of 800 pc) we find differences in the resulting velocity scatters of the test particles of less than 5 km/s. As seen in the table, the 
differences between the resulting velocity scatter when, in our spherical representation, changing from an analytical calculation of forces to the numerical interpolation are at the most 10\% of the 
scatter as such, and not greater than what must be expected from the limited statistics. Obviously, the major change of velocity scatter in $U$ and $V$ velocities results
from the spiral arms as such. From these different experiments we conclude that the
numerical uncertainties in the calculation of forces and orbits are of little significance. 
\begin{table}[tab:check_spheroid]
\caption{A test of the numerical errors when calculating forces from spiral-arms on test particles in our model simulations.  
Results when calculating the forces from  {\it spherical} spiral-arm elements,
by analytical calculation and by numerical interpolation following our standard procedure, respectively, are shown. In the bottom row the results of neglecting the spiral arms
altogether are displayed. $N(GMC)=10,000$,   Dispersions in $U$, $V$ and $W$ velocities in km/s are given.
The fraction of the test particles ending with a distance greater than 400 pc from the Galactic plane is $< 0.5\%$ in all cases.}
  \centering
  \begin{tabular}[l]{lccc}
    \hline
   Model   & $\sigma_U$ &$\sigma_V$  & $\sigma_W$\\
   \hline 
     analytical    &  46 & 22 & 8.9\\
     interpol.  &  42 & 20& 9.1\\
     no spiral arms& 25 & 12 & 8.4\\
   \hline
  \end{tabular}
  \label{tab:check_spheroid}
\end{table}

A probably much more severe source of error in our treatment than the numerical inaccuracies is the assumption that both the spiral arms and the Galactic bar are 
assumed to be stationary (although turning with constant angular velocities); more transient phenomena
might induce different, and probably greater, momentum transfer to our test particles in the model.
\end{appendix}

\bibliographystyle{aa}
\bibliography{sample.bib}

\end{document}